%% file: fpcp11-template_v2.tex
\documentclass[12pt]{article}
\usepackage{graphicx}

\newcommand\pubdate{\today}

\newcommand\pubnumber{}

\def\Title#1{\begin{center} {\Large #1 } \end{center}}
\def\Author#1{\begin{center}{ \sc #1} \end{center}}
\def\Address#1{\begin{center}{ \it #1} \end{center}}

\newcommand\pubblock{\rightline{\begin{tabular}{l} \pubnumber\\
         \pubdate  \end{tabular}}}
\newenvironment{Abstract}{\begin{center}{\bf Abstract}\end{center} \bigskip \begin{quotation}  }{\end{quotation}}
\newenvironment{Presented}{\begin{quotation} \begin{center} 
             PRESENTED AT\end{center}\bigskip 
      \begin{center}\begin{large}}{\end{large}\end{center} \end{quotation}}

\input econfmacros.tex

\begin{document}
\begin{titlepage}
\pubblock

\vfill


\Title{Searches for Fourth Generation Fermions}
\vfill
\Author{A. Ivanov}  
\Address{Kansas State University, Manhattan, KS, 66502, U.S.A.}
\vfill


\begin{Abstract}
We present the results from searches for fourth generation fermions 
performed using data samples collected by the CDF~II and D0 Detectors
at the Fermilab Tevatron $p\bar{p}$ collider.
Many of these results represent the most stringent 95\% C.~L.~limits on masses 
of new fermions to-date.
\end{Abstract}

\vfill

\begin{Presented}
The Ninth International Conference on\\
Flavor Physics and CP Violation\\
(FPCP 2011)\\
Maale Hachamisha, Israel,  May 23--27, 2011
\end{Presented}
\vfill

\end{titlepage}
\def\thefootnote{\fnsymbol{footnote}}
\setcounter{footnote}{0}
%


\section{Introduction}

A fourth chiral generation of massive fermions 
with the same quantum numbers as the known fermions 
is one of the simplest extensions of the SM with three generations.
The fourth generation is predicted in a number of
theories~\cite{democratic, gut, he}, and although historically have been 
considered disfavored,  
stands in agreement with electroweak precision data~\cite{he,okun,kribs}.

To avoid $Z\to \nu\bar{\nu}$ constraint from LEP~I  
a fourth generation neutrino $\nu_4$ must be heavy: $m(\nu_4) > m_Z/2$, 
where $m_Z$ is the mass of $Z$ boson, and to avoid LEP~II bounds 
a fourth generation charged lepton $\ell_4$ must have $m(\ell_4) >  101$  GeV/c$^2$. 
At the same time due to sizeable radiative corrections masses of fourth generation 
fermions cannot be much higher the current lower bounds and masses of 
new heavy quarks $\tprime$ and $b^\prime$ should be in the range of a few hundred 
 GeV/c$^2$~\cite{kribs}. 

In the four-generation model
the present bounds on the Higgs are relaxed: the Higgs 
mass could be as large as 1 TeV/$c^2$~\cite{kribs, frampton, ewk}. 
Furthermore, the CP violation 
is significantly enhanced to the magnitude
that might account for
the baryon asymmetry 
in the Universe~\cite{hou}.
Additional chiral fermion families can also be 
accommodated in supersymmetric two-Higgs-doublet 
extensions of the SM 
with equivalent effect on the precision fit to the Higgs mass~\cite{he}.

Another possibility is 
heavy exotic quarks with 
vector couplings to the $W$ boson~\cite{vectorlike}
Contributions to radiative corrections from such quarks 
with mass M decouple as $1/M^2$ 
and easily evade all experimental constraints.

At the Tevatron $\ppbar$ collider 4-th generation chiral or vector-like
quarks can be either produced strongly in pairs or singly via electroweak production, 
where the latter can be enhanced for vector-like quarks.
In the following we present searches for both pair and single production of heavy quarks
performed by CDF and D0 Collaborations.

\section{Search for pair production $t'\bar{t'}$ with subsequent decays $t' \to Wb$ and $t' \to Wq$}

Due to preferred 
small mass splitting between fourth generation $\tprime$ and $b^\prime$ quarks,
$m(b^\prime)+m(W) > m(\tprime)$,  
$\tprime$ decays predominantly to $Wq$ ( a $W$ boson and a down-type quark 
$q = d, s, b$)~\cite{kribs, frampton}.
CDF analyzed 5.6 fb$^{-1}$ of $\ppbar$ collisions data~\cite{CDF-tprime} to search for both possibilities: 
$q$ is a light ($d, s$), or heavy flavor quark ($b$).
Analysis is performed using events characterized by a high-$p_T$ lepton, large missing transverse
energy $\met$ and at least four hadronic jets. The data events were collected by triggers that identify at least 
one high-$p_T$ electron or muon candidate or by a trigger requiring $\met$ plus jets.

For the $\tprime \to Wb$ search at least one of the jets was required to be identified as coming 
from a bottom quark ($b$-tagged) by a secondary vertex tagging algorithm.
For the $\tprime \to Wq$  search instead, several additional kinematic event selection criteria were applied to suppress QCD background 
contributions. The main SM backgrounds for these searches are $\ttbar$ and $W+$ light- and heavy-flavor jets productions,
that are modeled with \textsc{PYTHIA} and \textsc{ALPGEN} generators.

The $\tprime$ events and mass of $\tprime$ quark were fully reconstructed using a $\chi^2$-based fit technique, 
used in top quark mass measurement analyses. The search for $\tprime$ signal is performed by employing a 2D 
binned likelihood fit for the mass of $\tprime$ quark ($M_{reco}$) and the $H_T = \sum_{jets} E_T + E_{T,\ell} + \met$, 
a scalar sum of transverse energies for all objects in the event, which also serves as a good discriminator between SM 
and new physics signal. To further enhance a sensitivity for $\tprime$, the events were split into four different sub-samples
based on the number of jets ( 4 or $\geq 5$) and based on good or poor mass reconstruction $\chi^2$.

\begin{figure}[htbp]
\centering
\includegraphics[width=0.49\textwidth]{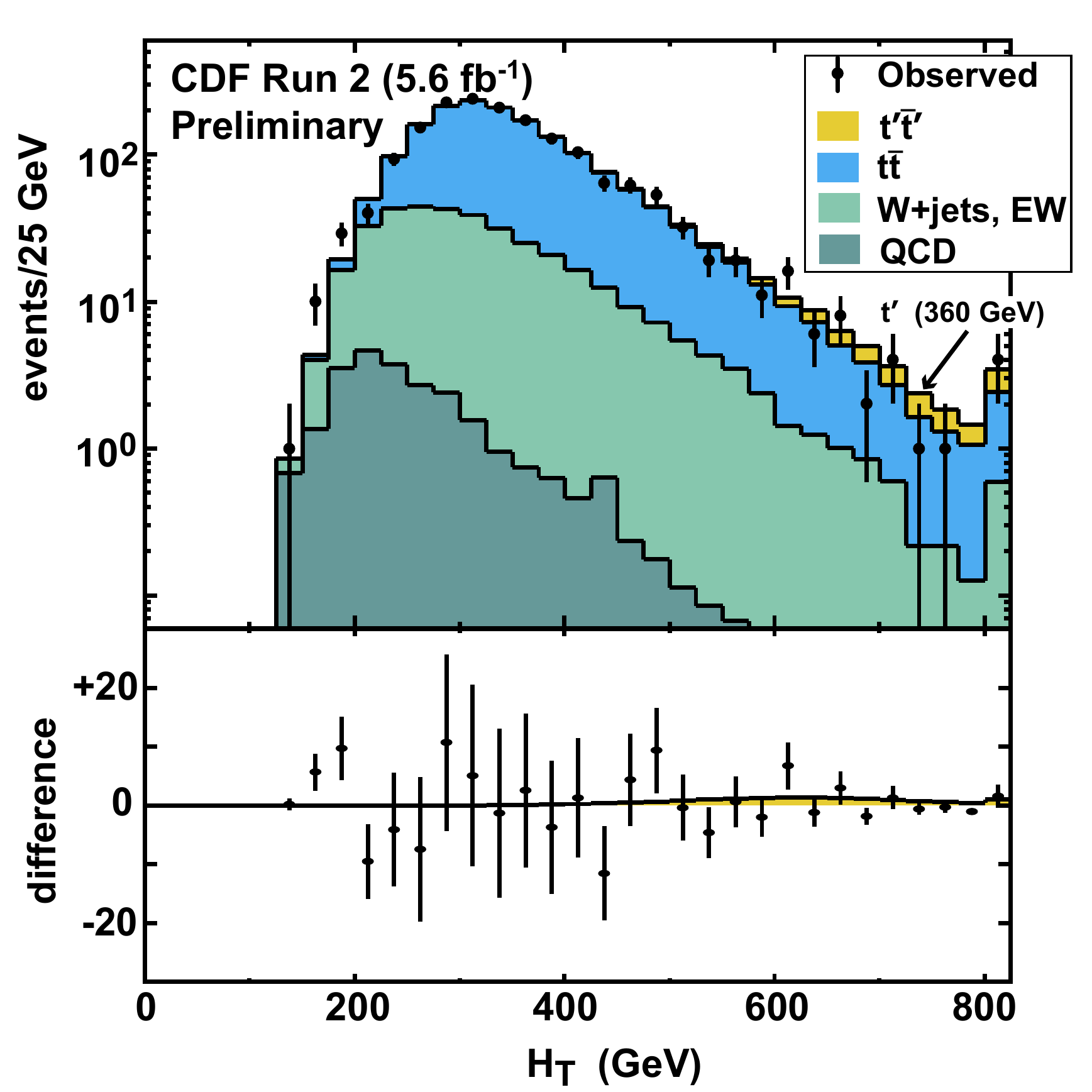}
\includegraphics[width=0.49\textwidth]{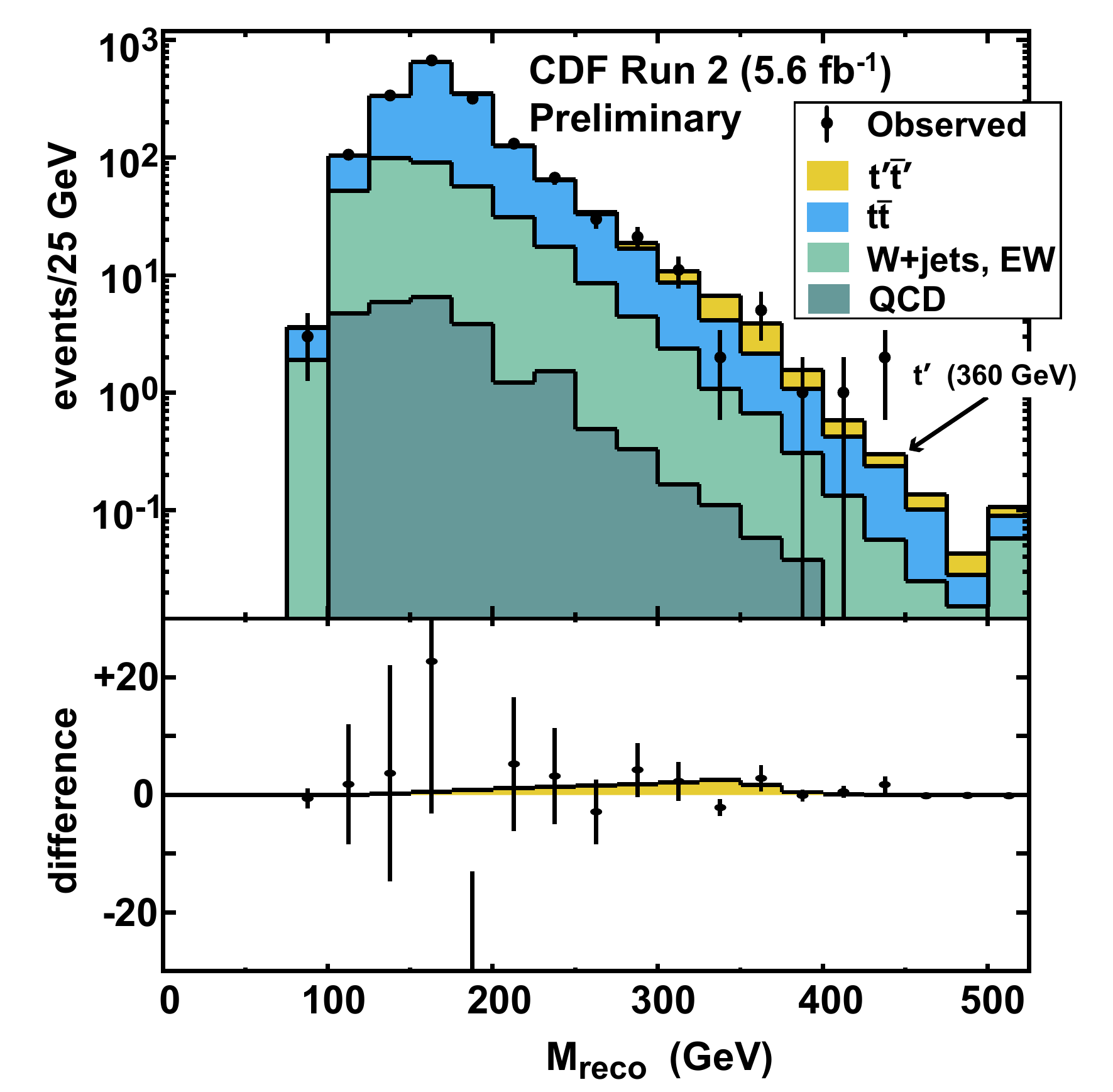}
\caption{Distributions of $H_T$ and $M_{reco}$. The contribution from $\tprime$ events shown on the plot 
corresponds to $\tprime$ mass of 360 GeV/c$^2$, assuming strong production mechanism.}
\label{fig:HT-m360}
\end{figure}

\begin{figure}[htbp]
\centering
\includegraphics[width=0.6\textwidth]{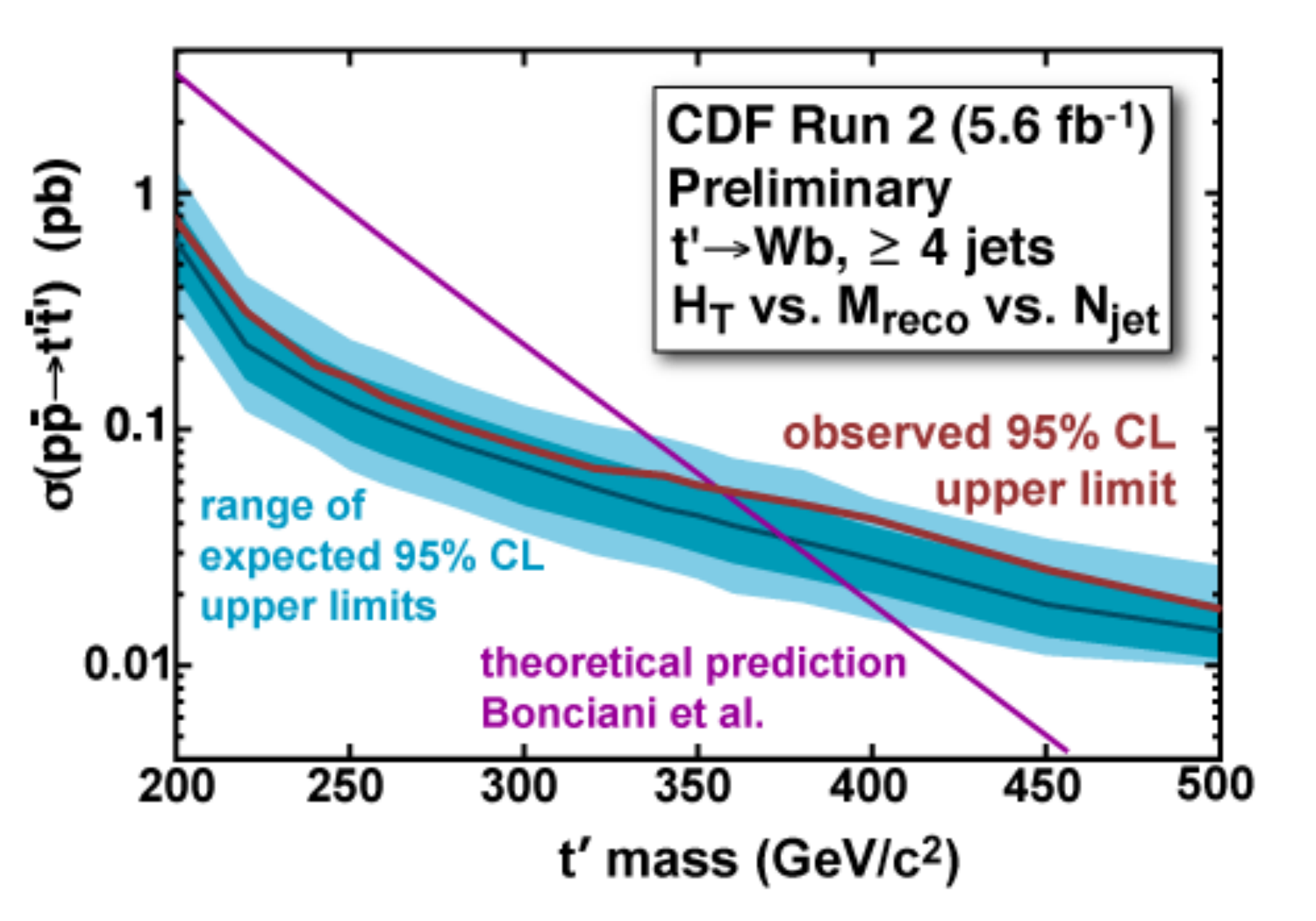}
\caption{ Observed and expected 95\% C.L. upper limits as a function of the mass of the $\tprime$ quark, for a $\tprime$ decaying to $Wb$, obtained in CDF analysis.}
\label{fig:limits}
\end{figure}

The kinematic distributions of $H_T$ and $M_{reco}$ for the $\tprime \to Wb$ analysis are shown in Fig.~\ref{fig:HT-m360}.
The analysis fit on the data shows no significant excess due to $\tprime \bar{\tprime}$ production.
The observed 95\% C.L. limits along with expected limits bands are derived using Bayesian technique and are
shown in Fig.~\ref{fig:limits}. 
Assuming strong production mechanism we exclude $\tprime$ quark with mass below 358 GeV/c$^2$ for $B(\tprime \to Wb) = 100\%$, 
and below 340 GeV/c$^2$ for $B(\tprime \to Wq) = 100\%$, where $q$ is a light flavor quark, at 95\% C.L.

\begin{figure}[htbp]
\centering
\includegraphics[width=0.99\textwidth]{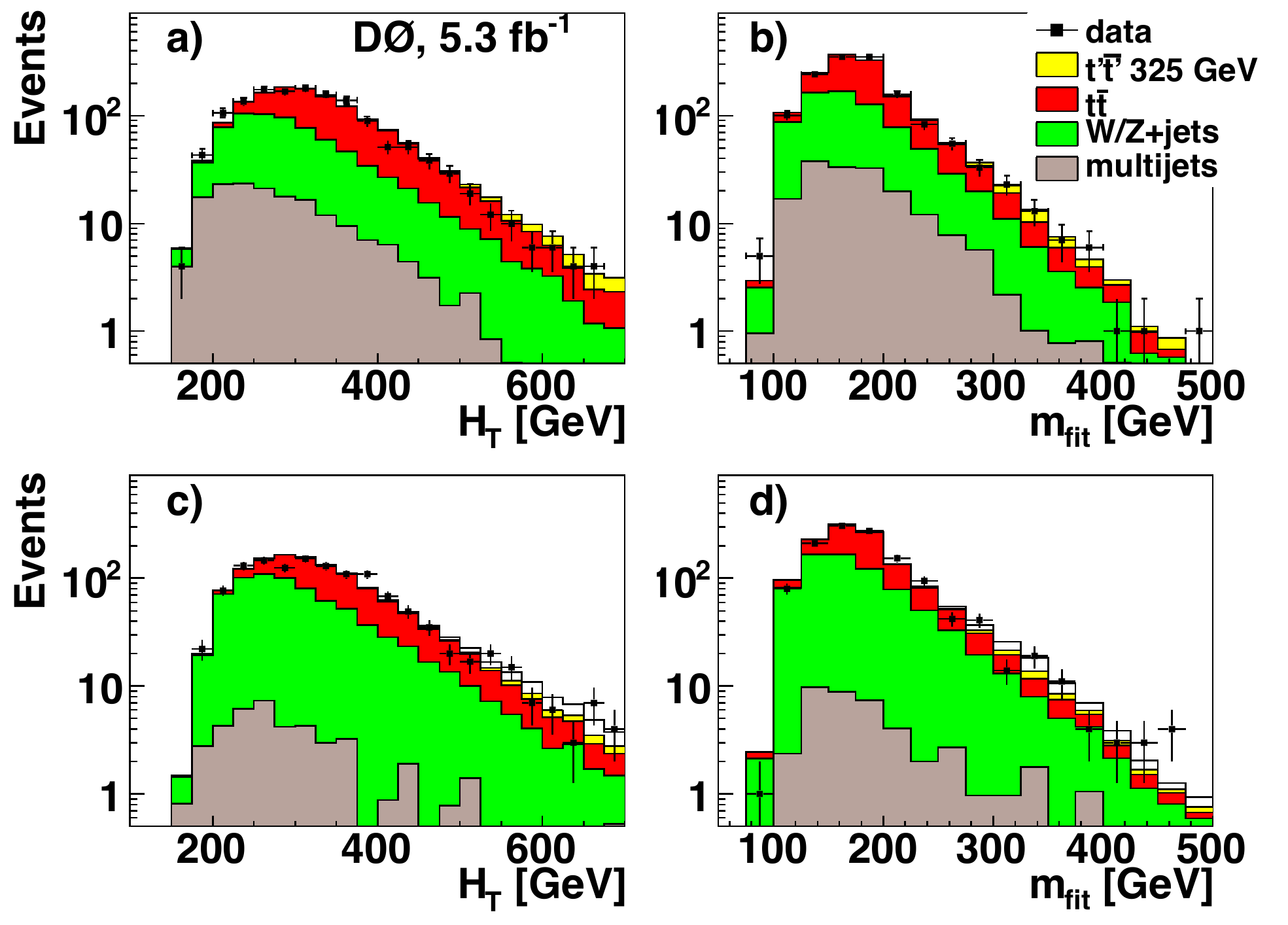}
\caption{Distributions of $H_T$ and $\tprime$ reconstructed mass for $e+$ jets (top) and $\mu+$ jets events (bottom).
}
\label{fig:T11HF02}
\end{figure}

\begin{figure}[htbp]
\centering
\includegraphics[width=0.49\textwidth]{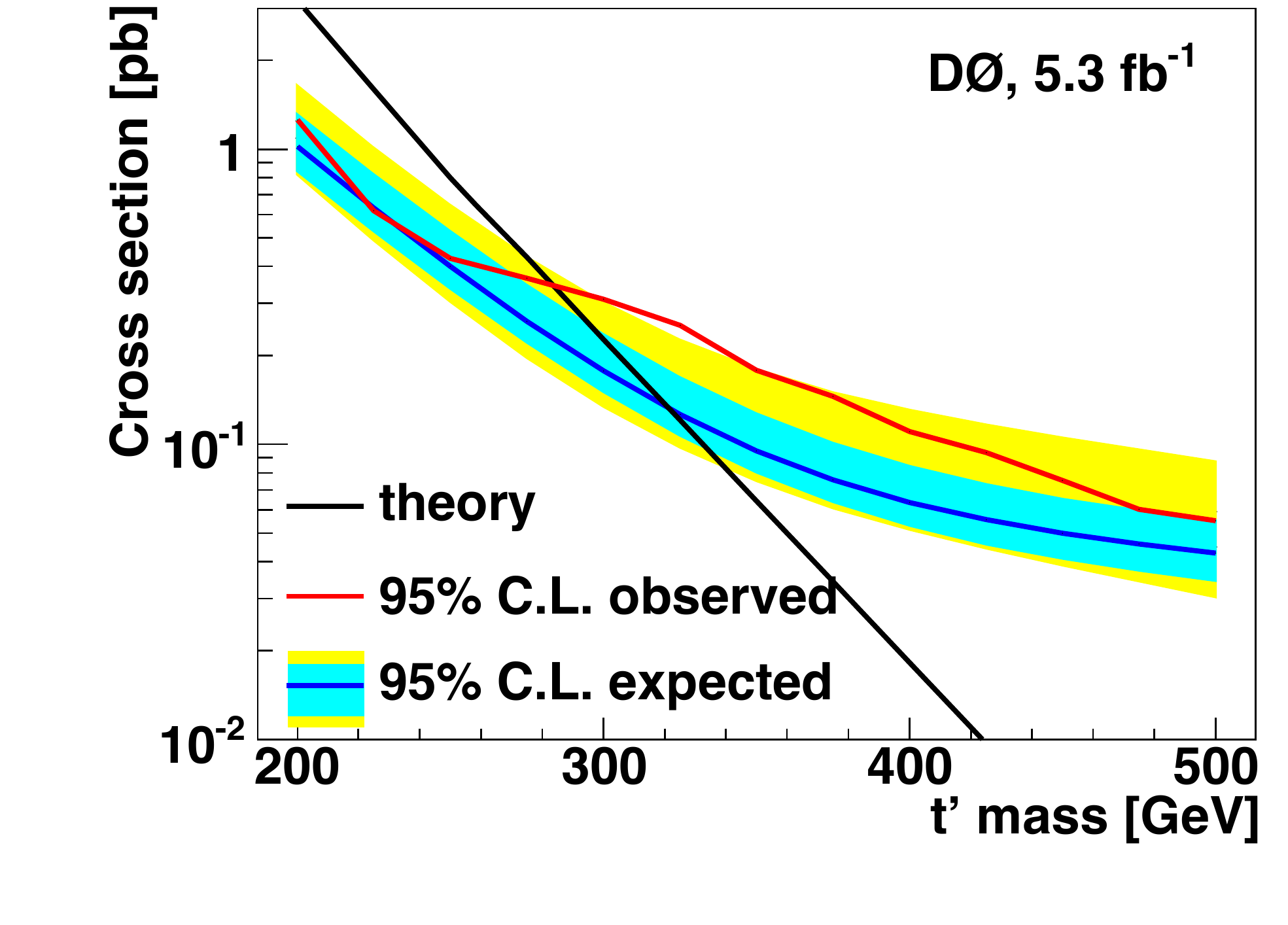}
\caption{Observed and expected upper limits on $\tprime \bar{\tprime}$ production as a function of $\tprime$ mass, obtained in D0 
analysis.}
\label{fig:T11HF04}
\end{figure}

Similar search for $\tprime \to Wq$ is done by D0 Collaboration~\cite{D0-tprime}, with the fit for $e+$ and $\mu+$ jets events performed separately.
The kinematic distributions of $H_T$ and $M_{reco}$ are shown in Fig.~\ref{fig:T11HF02}.
The data show no evidence for $\tprime$, and the 95\% C.L. observed and expected limits are presented in Fig.~\ref{fig:T11HF04}.

\section{Search for pair production $b'\bar{b'}$ with subsequent decay $b' \to tW$ }

If the coupling of down-type fourth-generation quark $\bprime$ to light quarks is small, then $\bprime$ decays 
exclusively to $tW$. The limits on $\bprime \to Wq$ mode can be assessed from the $\tprime$ analysis.
CDF analyzed 4.8 fb$^{-1}$ of data considering 
the mode $b'\bar{b'} \to WtW\bar{t} \to WWbWW\bar{b}$ in which one $W$ boson decays 
leptonically~\cite{CDF-bprime}. Production and decay of $\bprime$ pairs would appear as events with a charged lepton and 
missing transverse energy $\met$ from leptonically decaying $W$, and a large number of jets from the 
two $b$ quarks and the hadronic decays of the other three $W$ bosons. Selected events are those with at least 
five jets, with at least one of them identified as due to $b$ quark decay.
SM backgrounds are primarily due to $\ttbar+$ and $W+$ jets, modeled with \textsc{MADGRAPH} and \textsc{ALPGEN} respectively, 
and interfaced with \textsc{PYTHIA}.

A $\bprime$ signal is separated from the SM background both in the number of jets and the $H_T$.
To take advantage of both of these characteristics, a variable ``Jet-$H_T$" is introduced equal to $H_T + 1000$ GeV $\times (N_{jets}-5)$, which is equivalent to a two-dimensional analysis in $N_{jets}$ and $H_T$.
The description of SM backgrounds is validated in low $H_T$ regions. The CDF data distribution of Jet-$H_T$ variable 
is shown in Fig.~\ref{fig:htnj_paper}. In events with $\geq 7$ jets a mild excess is observed at high-$H_T$ region, 
however the total number of events in the $\geq 7$ jets  is consistent with expectation.
The upper limits on $\bprime\bar{\bprime}$ production are derived using Feldman-Cousins likelihood-ratio ordering~\cite{FC},
and are presented in Fig.~\ref{fig:htnj_paper}.

CDF excludes a $\bprime$ quark with mass below 372 GeV/c$^2$ at 95\% C.L., assuming strong pair production mechanism
and exclusive decay to $tW$.

\begin{figure}[htbp]
\centering
\includegraphics[width=0.49\textwidth]{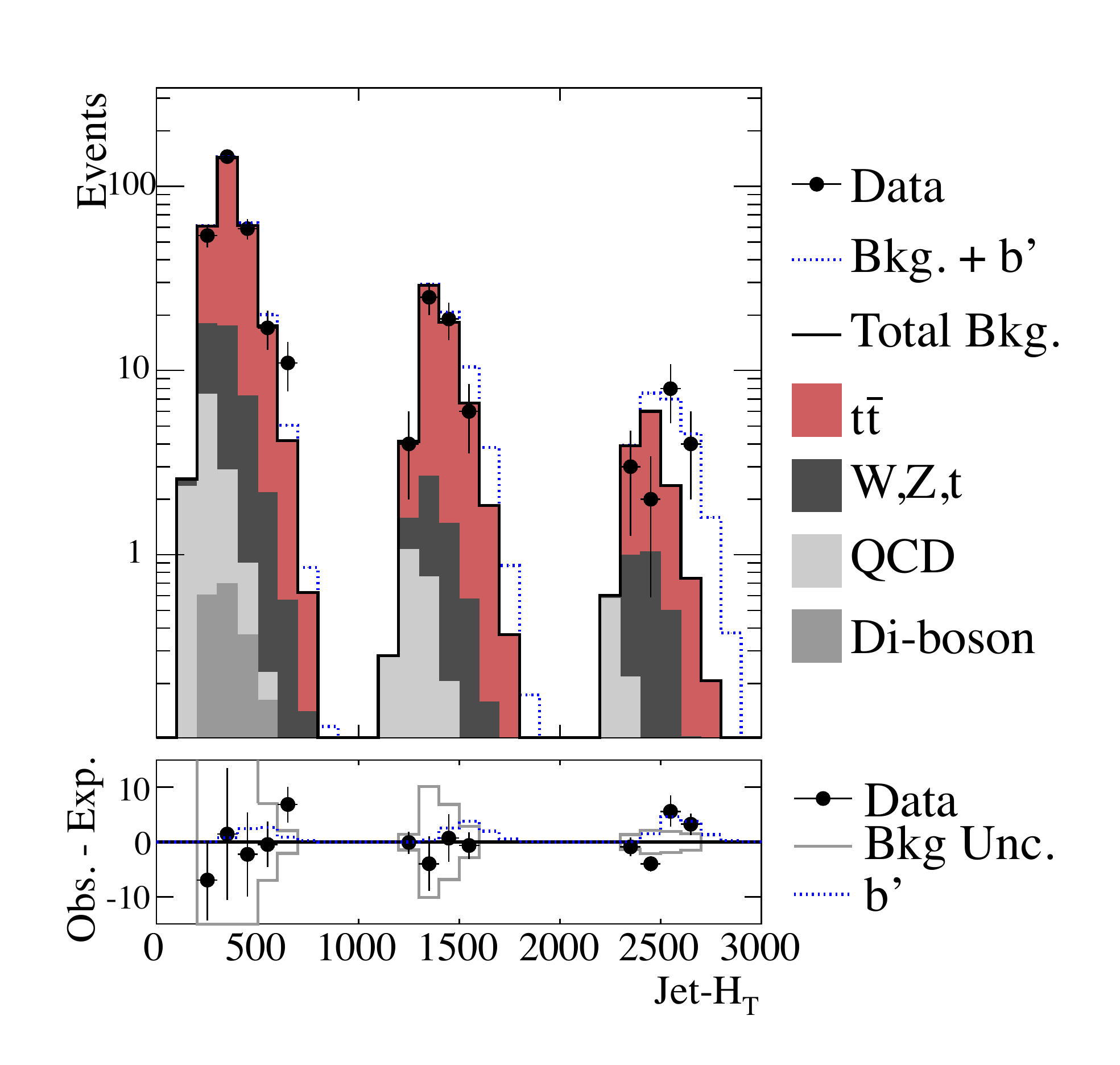}
\includegraphics[width=0.49\textwidth]{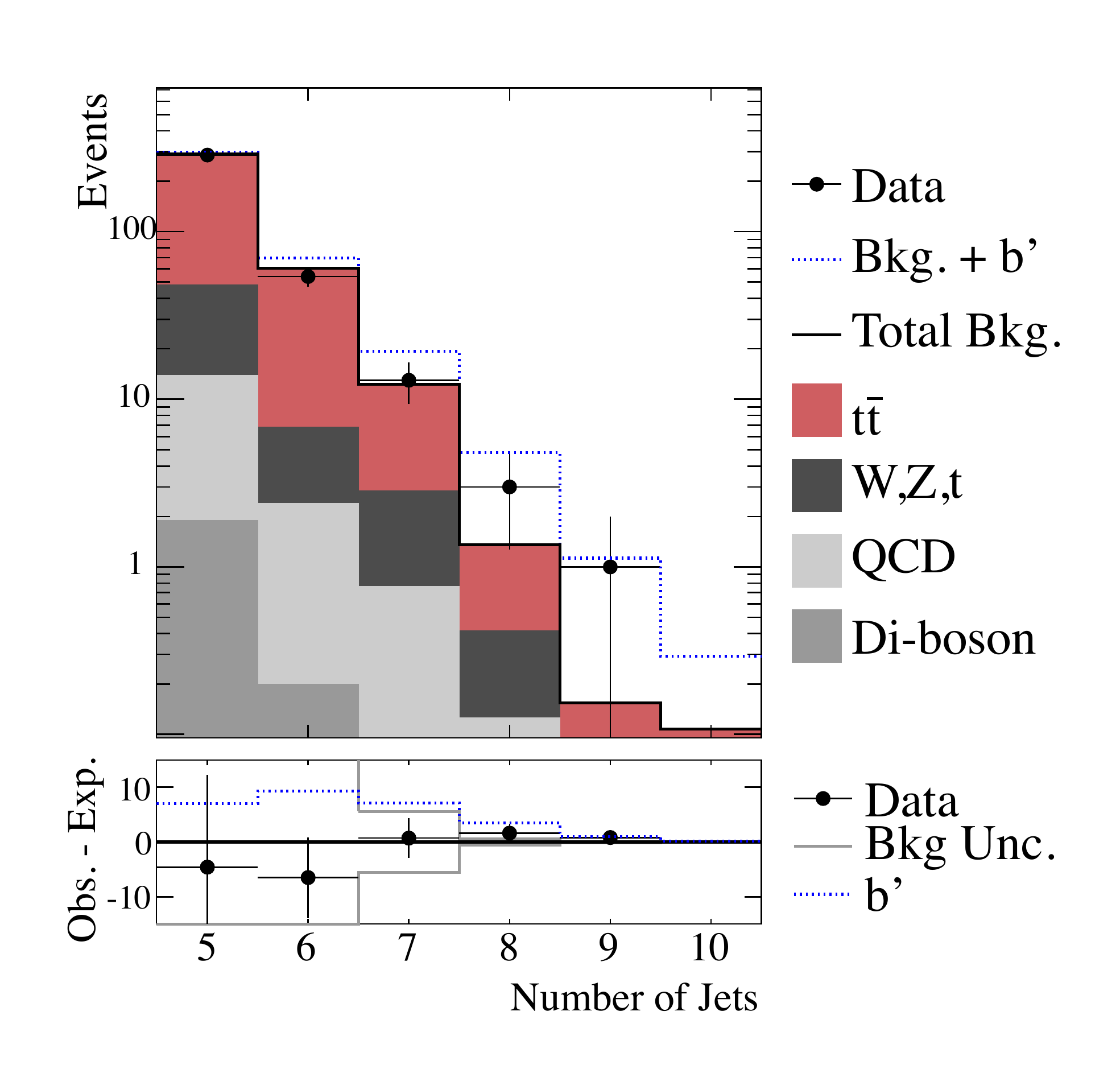}
\caption{The distributions of ``Jet-$H_T$" and jet multiplicity distributions in CDF $\bprime$ search.}
\label{fig:htnj_paper}
\end{figure}

\begin{figure}[htbp]
\centering
\includegraphics[width=0.6\textwidth]{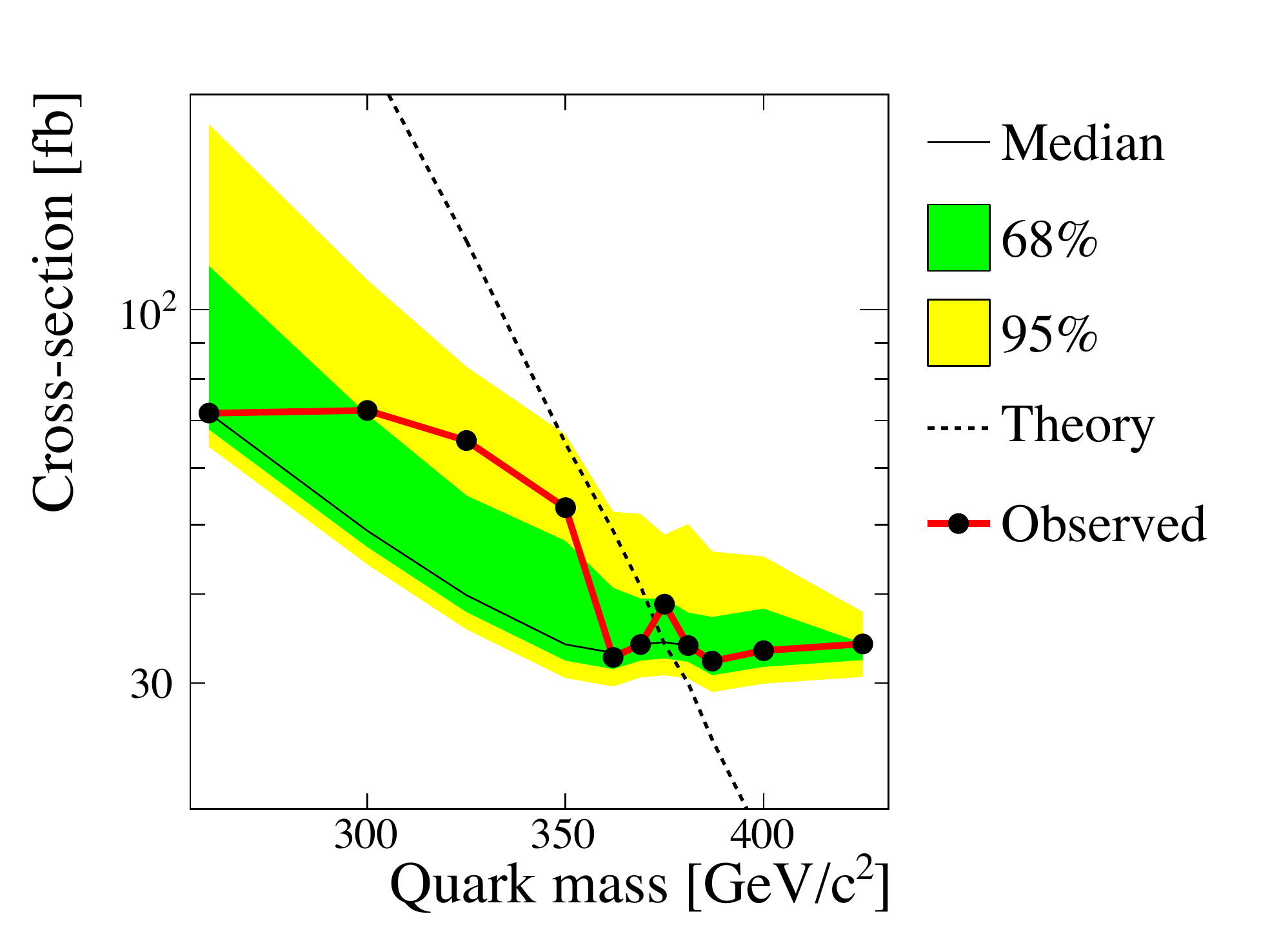}
\caption{The CDF upper limits on the $\bprime$ pair production, as a function of mass for $\bprime$ quark.}
\label{fig:htnj_paper}
\end{figure}

\section{Search for $t'\bar{t}$ with $t' \to th$ }

\begin{figure}[htbp]
\centering
\includegraphics[width=0.49\textwidth]{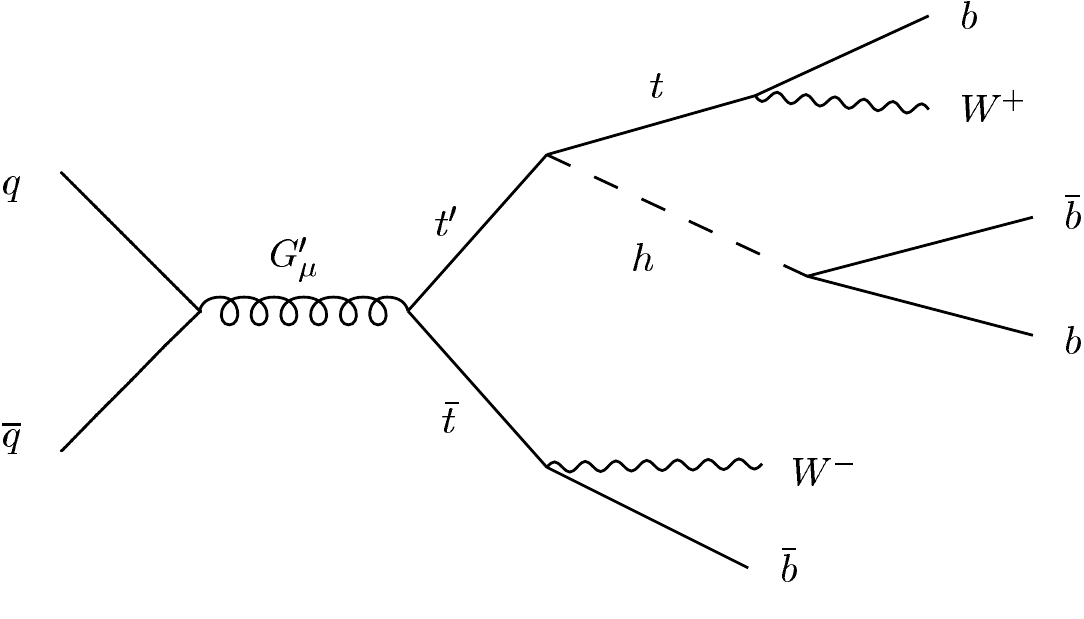}
\caption{Feynman diagram for axi-gluon production.}
\label{fig:H75F6}
\end{figure}

If the $\tprime$ quarks are produced via a new production mechanism, such as 
via new massive color-octet vector boson ($G'$) exchange, the production cross 
section can be substantially higher than the one from QCD~\cite{axigluon}. 
In these models it is possible that $G' \to t'\bar{t}$ with $\tprime \to th \to t b\bar{b}$~(see Fig.~\ref{fig:H75F6}).
This leads to event signatures with a large number of jets, three of which are expected 
to be from $b$-quarks.

\begin{figure}[htbp]
\centering
\includegraphics[width=0.49\textwidth]{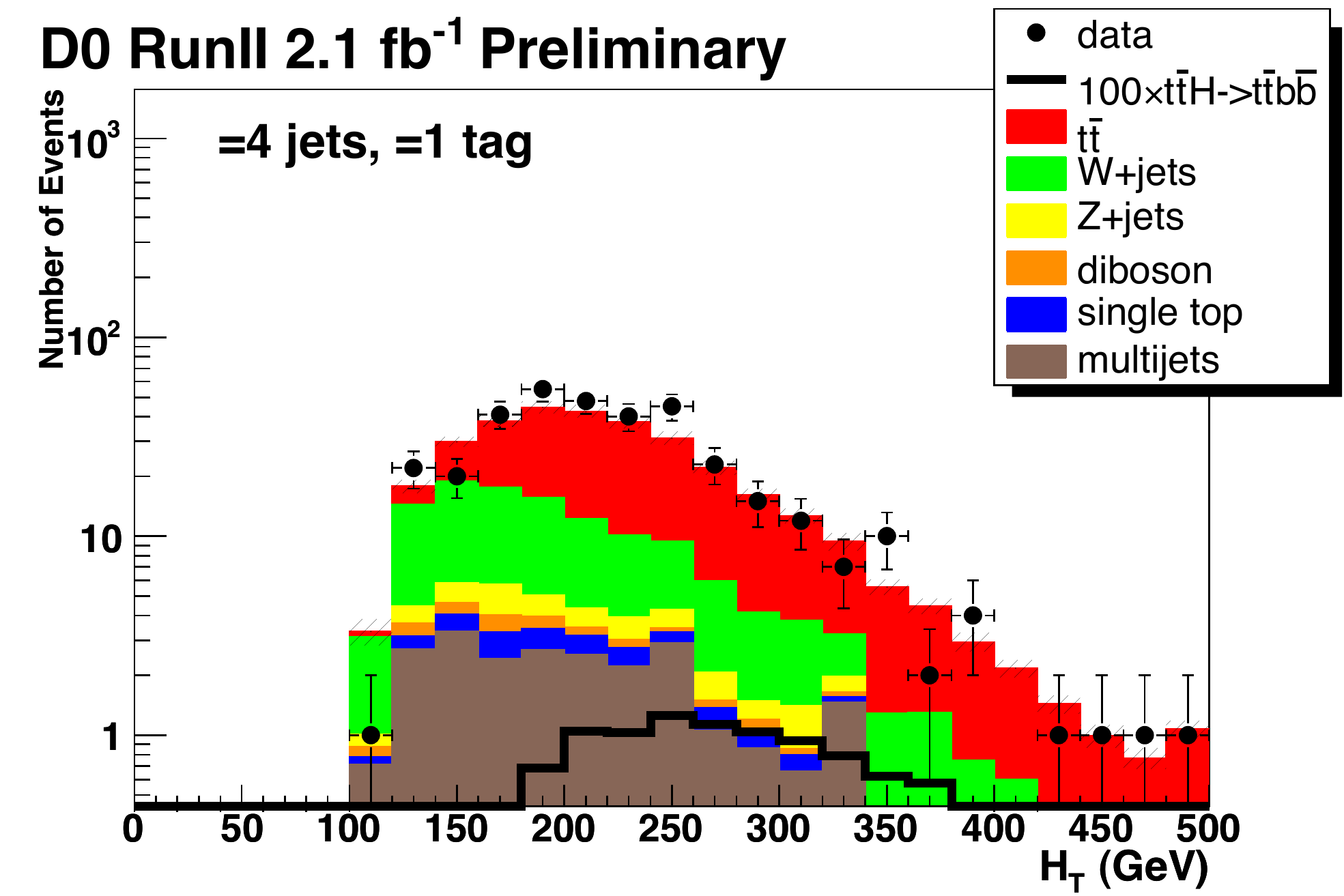}
\includegraphics[width=0.49\textwidth]{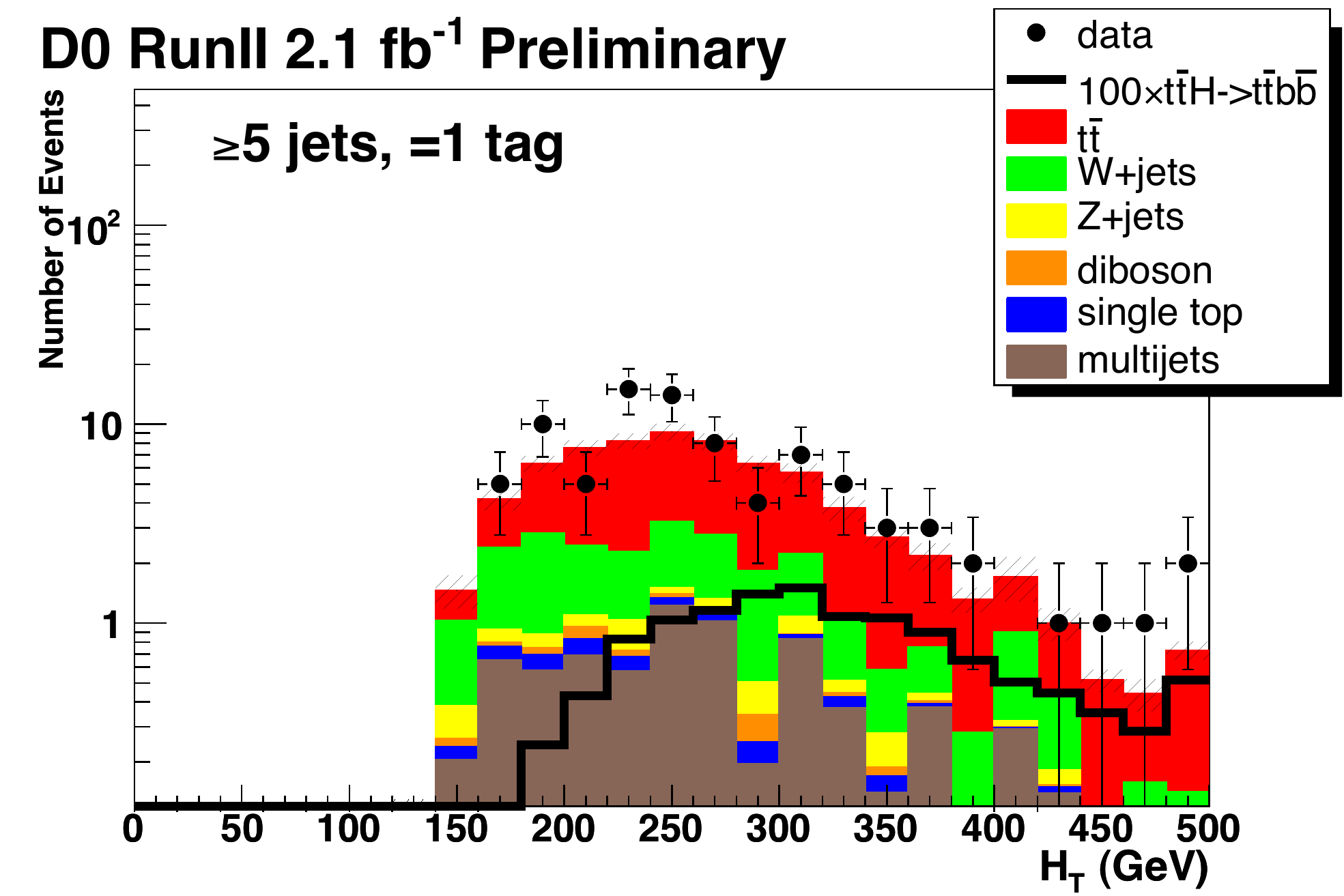}
\includegraphics[width=0.49\textwidth]{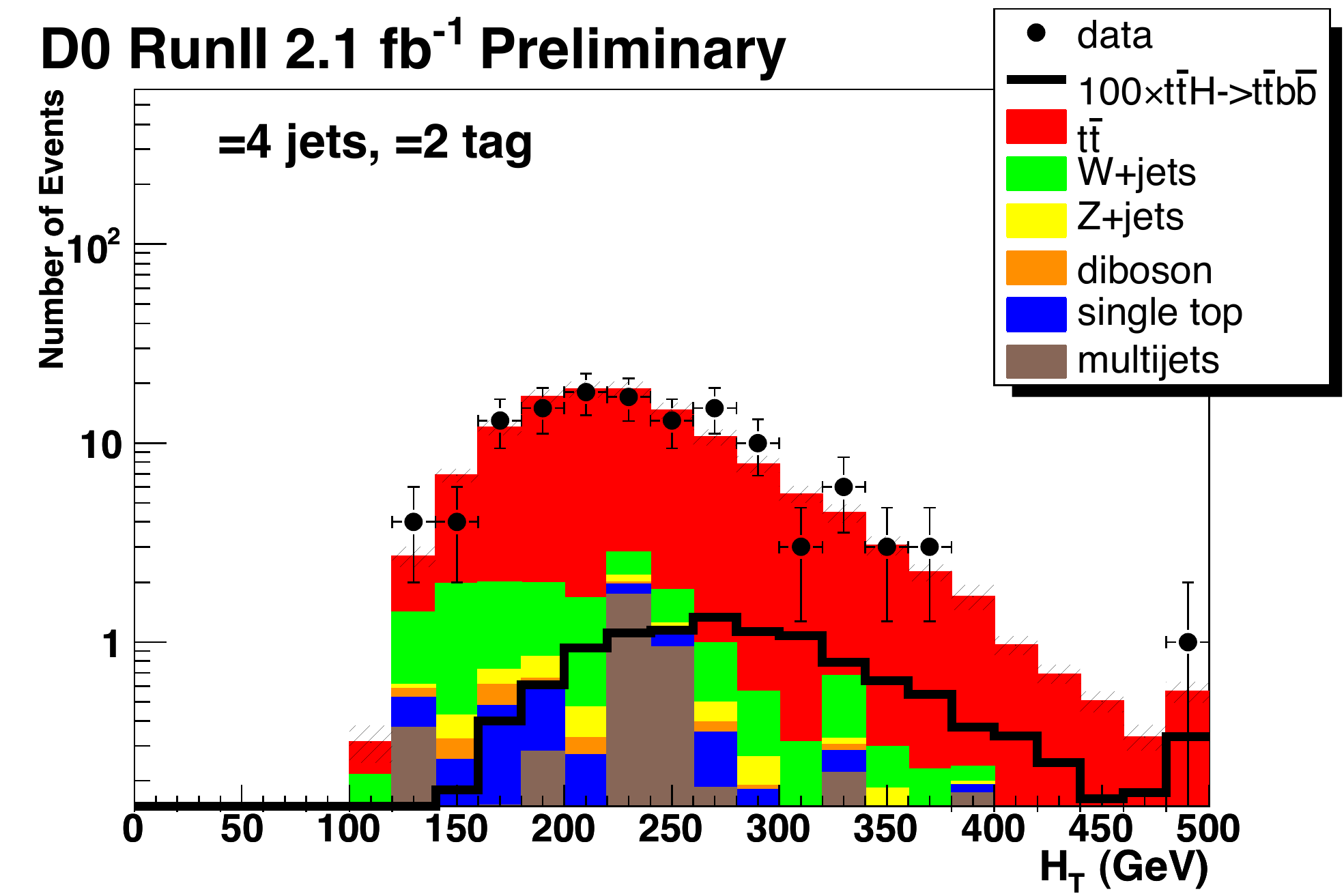}
\includegraphics[width=0.49\textwidth]{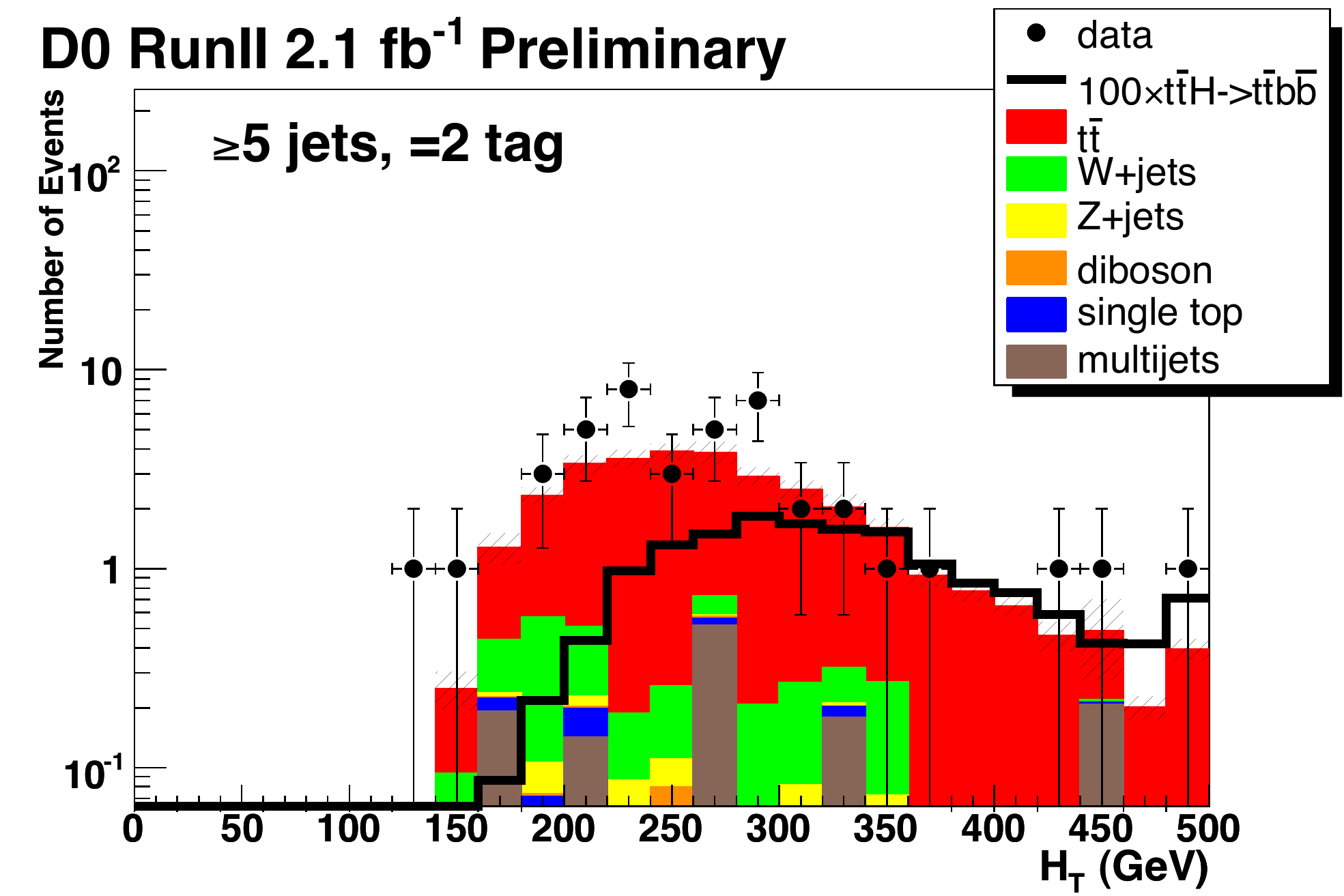}
\includegraphics[width=0.49\textwidth]{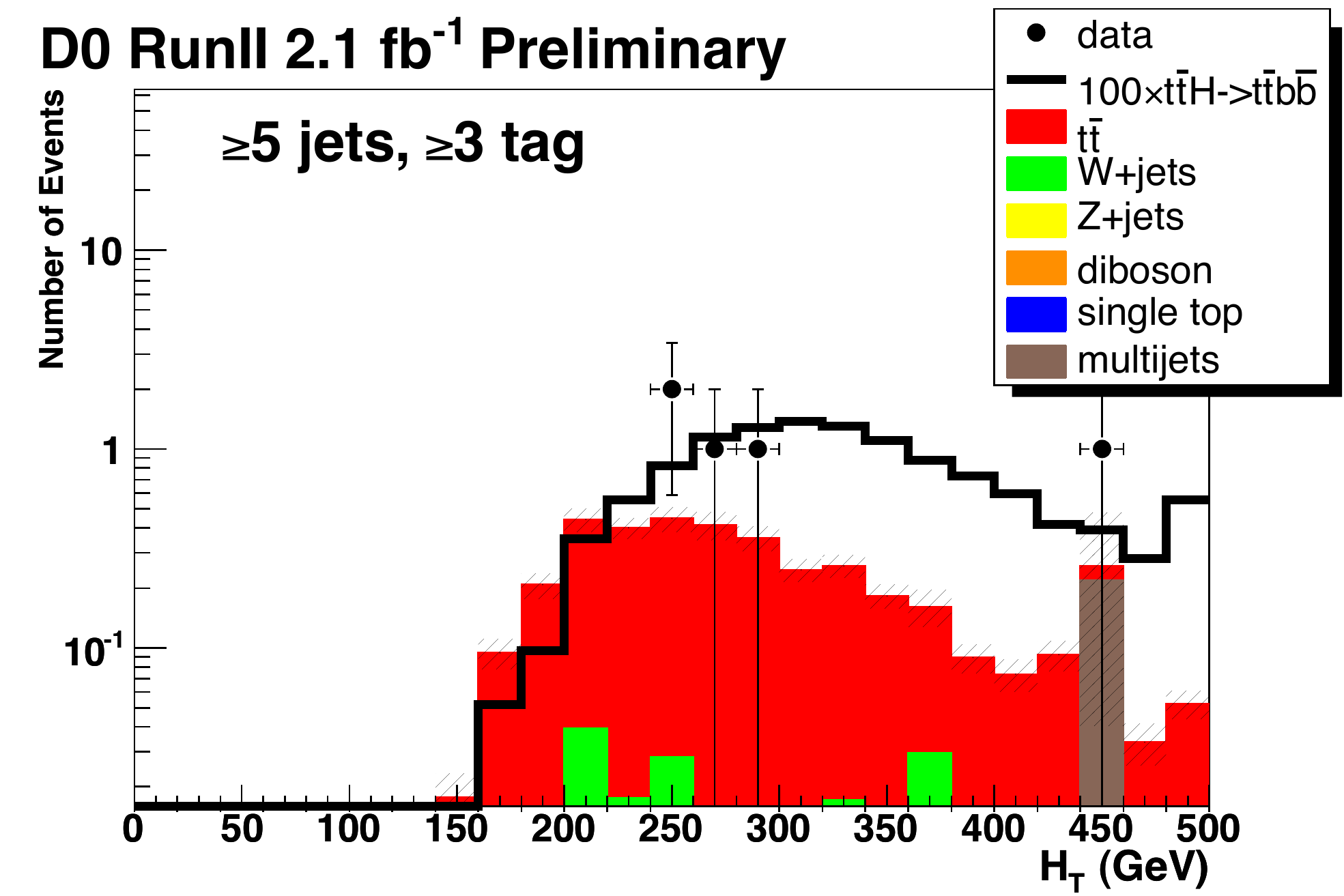}
\includegraphics[width=0.49\textwidth]{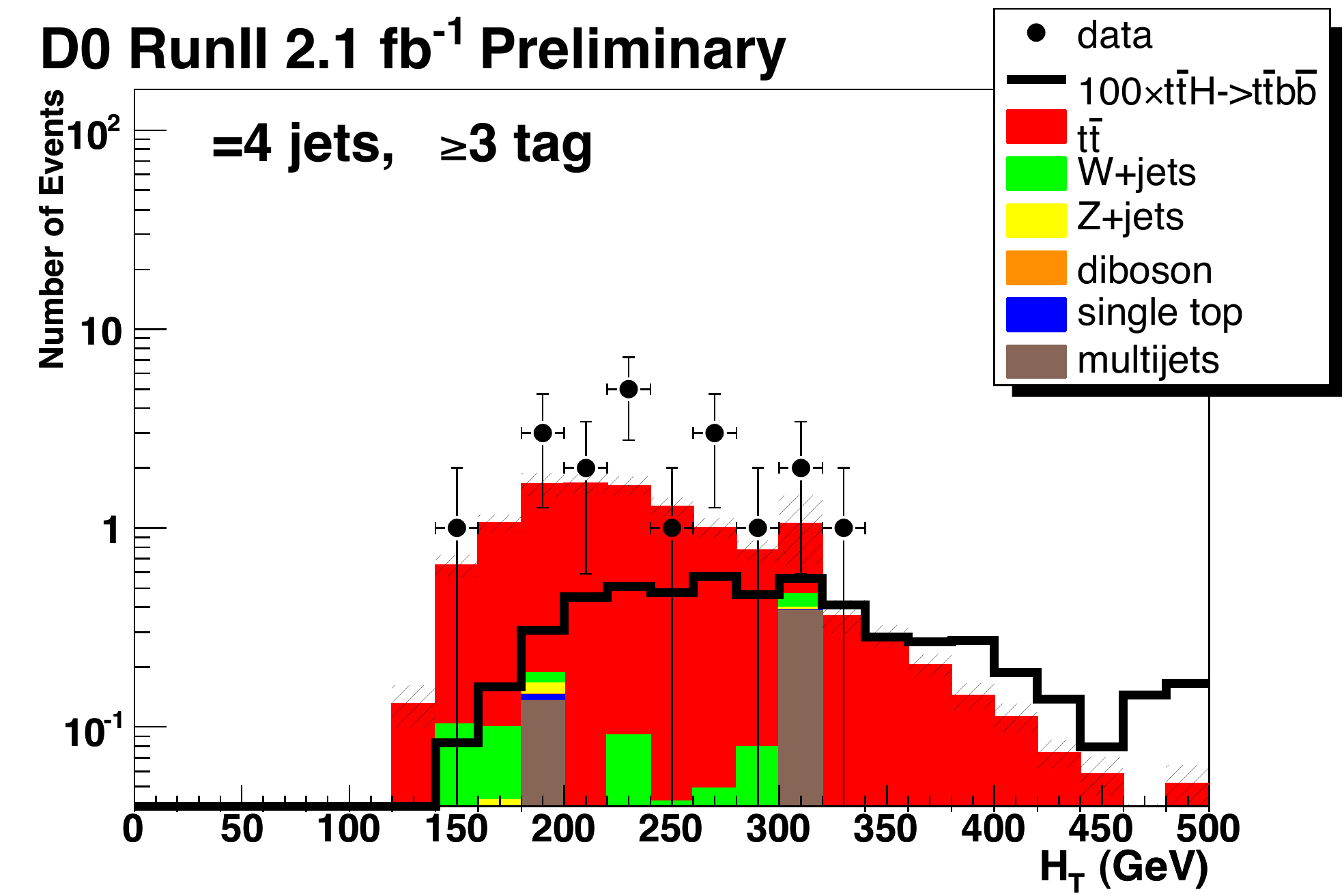}
\caption{$H_T$ distribution in various sub-samples of events, used in D0 search for $\tprime t$.}
\label{fig:H75F3a}
\end{figure}

D0 performed a search for $G' \to t'\bar{t} \to t\bar{t} h \to t\bar{t} b\bar{b}$ by using events with a least 
one high-$p_T$ lepton, large missing transverse energy and at least 4 jets, with at least one of them 
identified as a $b$-jet. The events are further split depending on the number of jets ( 4 or $\geq 5$ ), 
and the number of $b$-tagged jets ( 1, 2, or $\geq 3$ ). Next, the simultaneous fit to the $H_T$ distribution 
in all of these categories performed. The main background is from $t\bar{t}+$ jets, that is modeled with 
ALPGEN + PYTHIA, with a conservative 50\% systematic uncertainty assigned to the process $t\bar{t} b\bar{b}$.
The kinematic distributions of the $H_T$ variable for different number of jets and tags are shown in Fig.~\ref{fig:H75F3a}.

\begin{figure}[htbp]
\centering
\includegraphics[width=0.49\textwidth]{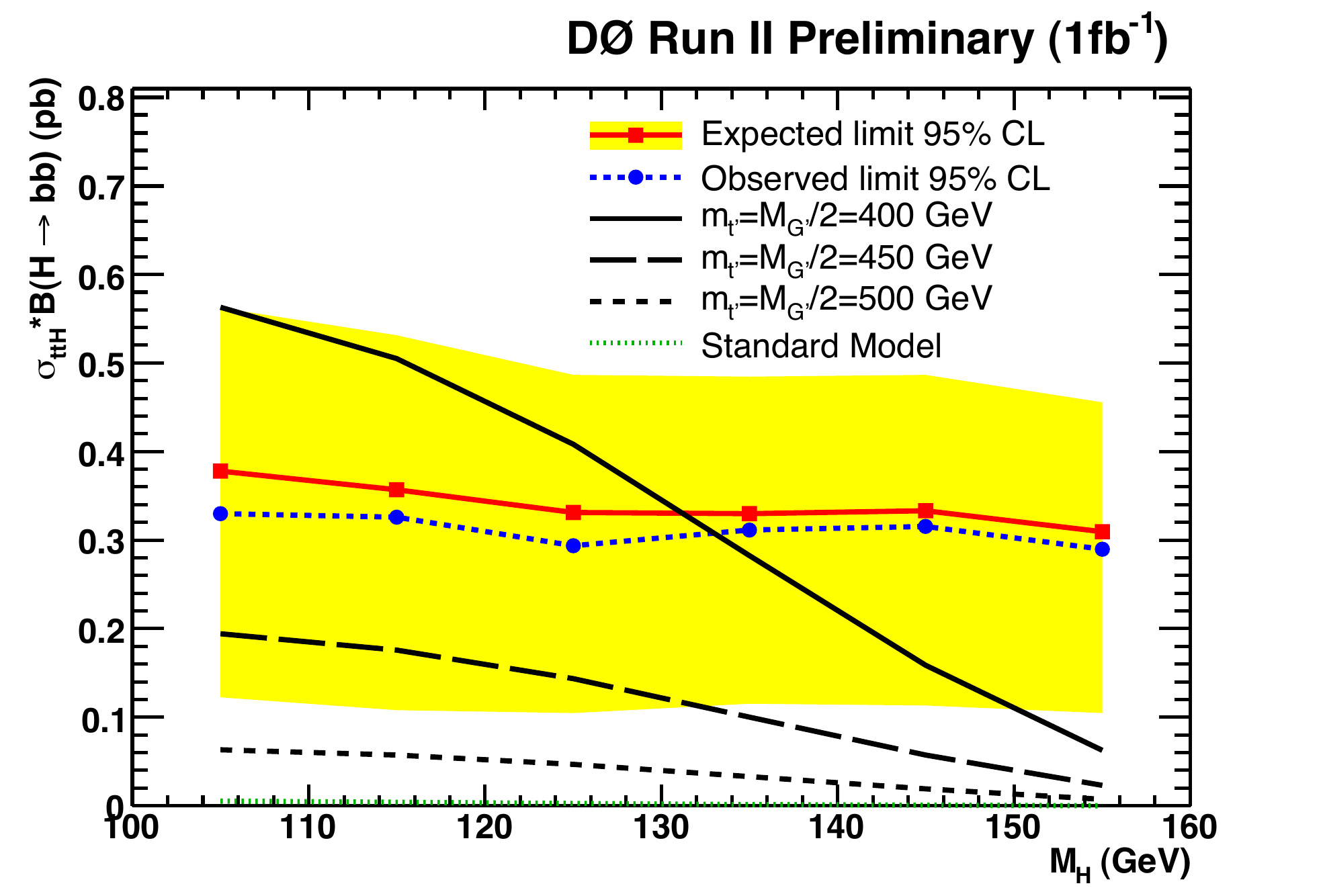}
\includegraphics[width=0.49\textwidth]{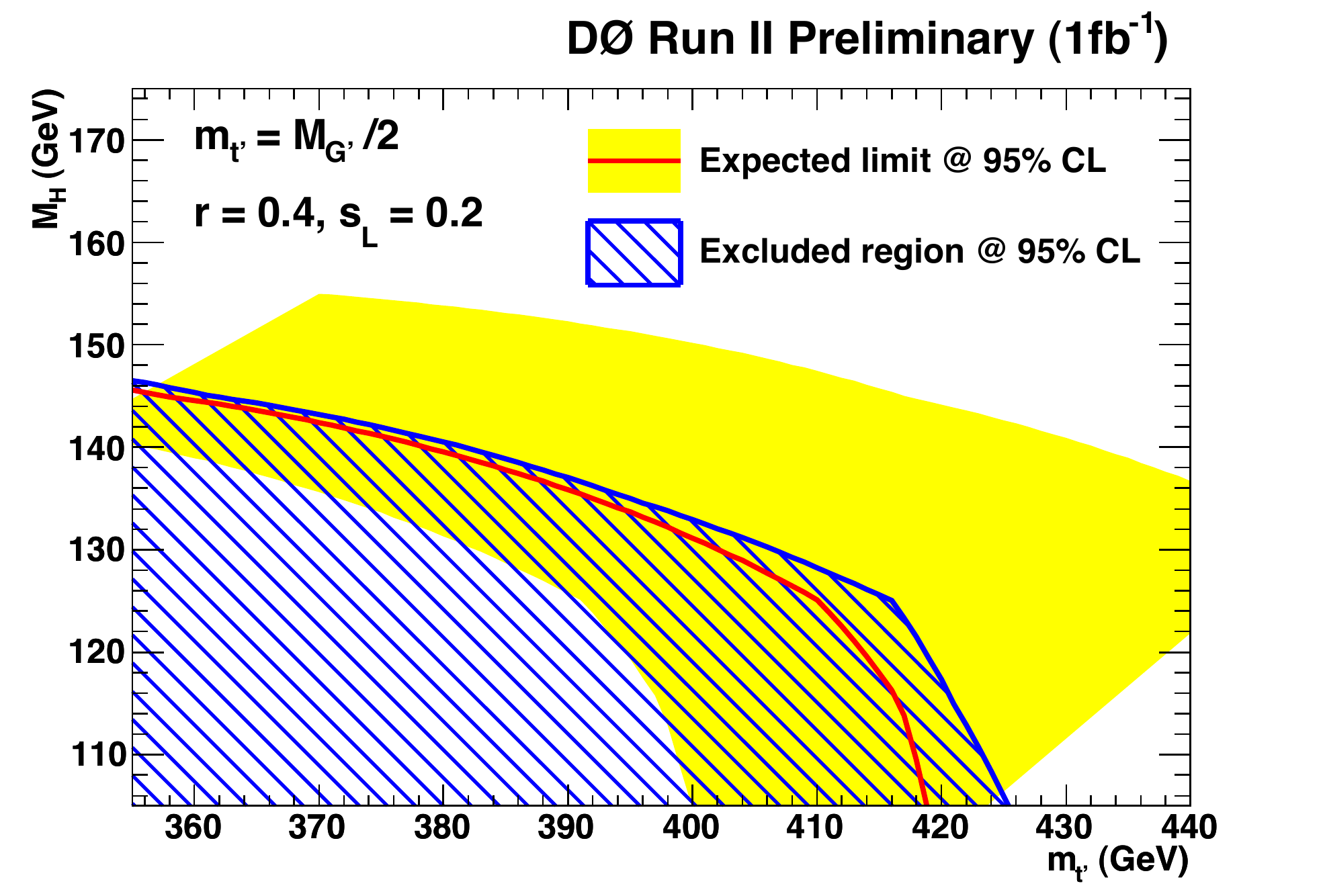}
\caption{{\bf LEFT:} The 95\% C.L. upper limit on the cross section times SM branching ratio of $H\to b\bar{b}$, as a function of the 
SM Higgs mass. {\bf RIGHT: }Excluded region as a function of the Higgs mass and the $\tprime$ mass, assuming $m_\tprime = M_{G'}/2, 
r = 0.4$ and $s_L$ .}
\label{fig:H75F7}
\end{figure}

The upper limits on $G'$ production are set using a modified frequentist approach ($CL_s$)~\cite{cls},
and shown in Fig.~\ref{fig:H75F7} as a function of Higgs mass for different masses of $\tprime$ quark. 
This can be interpreted as a 2D-limit in the phase space ($m_H, m_\tprime$) for a given coupling 
strength $r$ and a mixing angle $\sin \theta_L \equiv s_L$ between the top and $\tprime$ quark. 
The excluded region is shown in Fig.~\ref{fig:H75F7}.


\section{Search for single new heavy quarks $Q \to Wq$ and $Q \to Zq$}

\begin{figure}[htbp]
\centering
\includegraphics[width=0.49\textwidth]{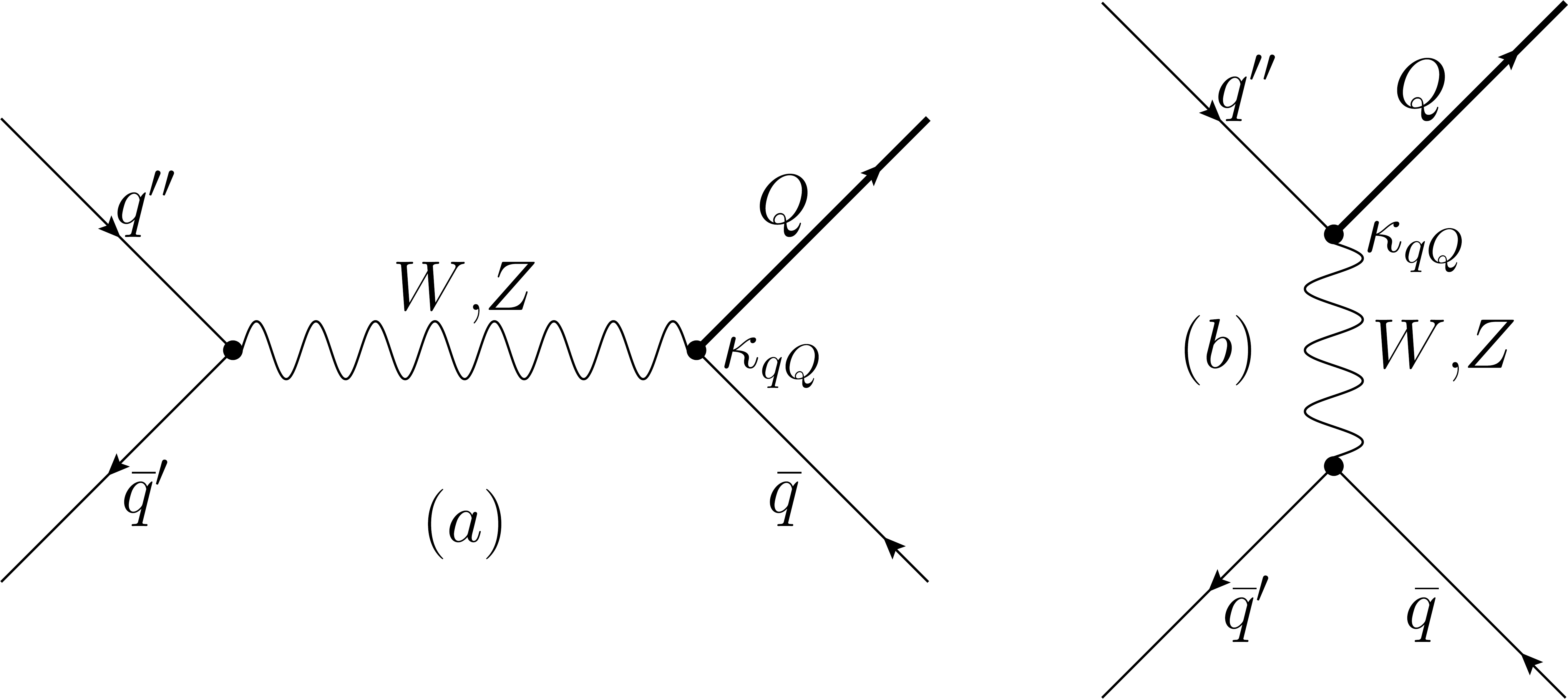}
\caption{Feynmann diagrams for single vector-like quark production.}
\label{fig:N10IF01}
\end{figure}

Single production of heavy quarks can occur due to diagrams shown in Fig.~\ref{fig:N10IF01}.
For vector-like quarks the production cross section can be enhanced with respect to SM-like 
fourth generation quarks, if the coupling strength $\kappa_{qQ}$ to SM quarks is sufficiently large~\cite{vector-like}.
Depending on the couplings to $u$ and $d$ quarks the new heavy quarks can decay to either $Wq$ or $Zq$.

D0 performed a search for vector-like quarks in both $(W \to \ell \nu) + $ jets and $(Z \to \ell \ell)$ + jets channels in 5.4 fb$^{-1}$ of data~\cite{D0-single}, 
using events with either single lepton + missing transverse energy ($W$), or events with exactly two electrons or muons forming 
an invariant mass consistent with that of a $Z$ boson. In addition, in both cases at least two high-$E_T$ jets are required to be present 
in the event. 

In the single lepton channel the highest-$E_T$ jet expected to be from vector-like quark decay, and required to have $E_T > 100$ GeV. 
The second-highest $E_T$ jet is expected to be from the SM quark produced in association with the vector-like quark. It is expected to be in one of the forward regions of the detector with direction strongly correlated with the charge of the produced vector-like quark, and thus also with the charge of the lepton from its decay. 
It is therefore required that $Q_\ell \times \eta_{j_2} > 0$, where $Q_\ell$ is the lepton charge and $\eta_{j_2}$ is the pseudorapidity
of the second jet in the event.
Similarly, in the dilepton channel additional selection requirements characteristic of a heavy resonance decay to a $Z$ boson and a jet
are applied, such as the $p_T$ of the dilepton system to be greater than 100 GeV, and the leading jet $E_T$ to be above 100 GeV.

\begin{figure}[htbp]
\centering
\includegraphics[width=0.49\textwidth]{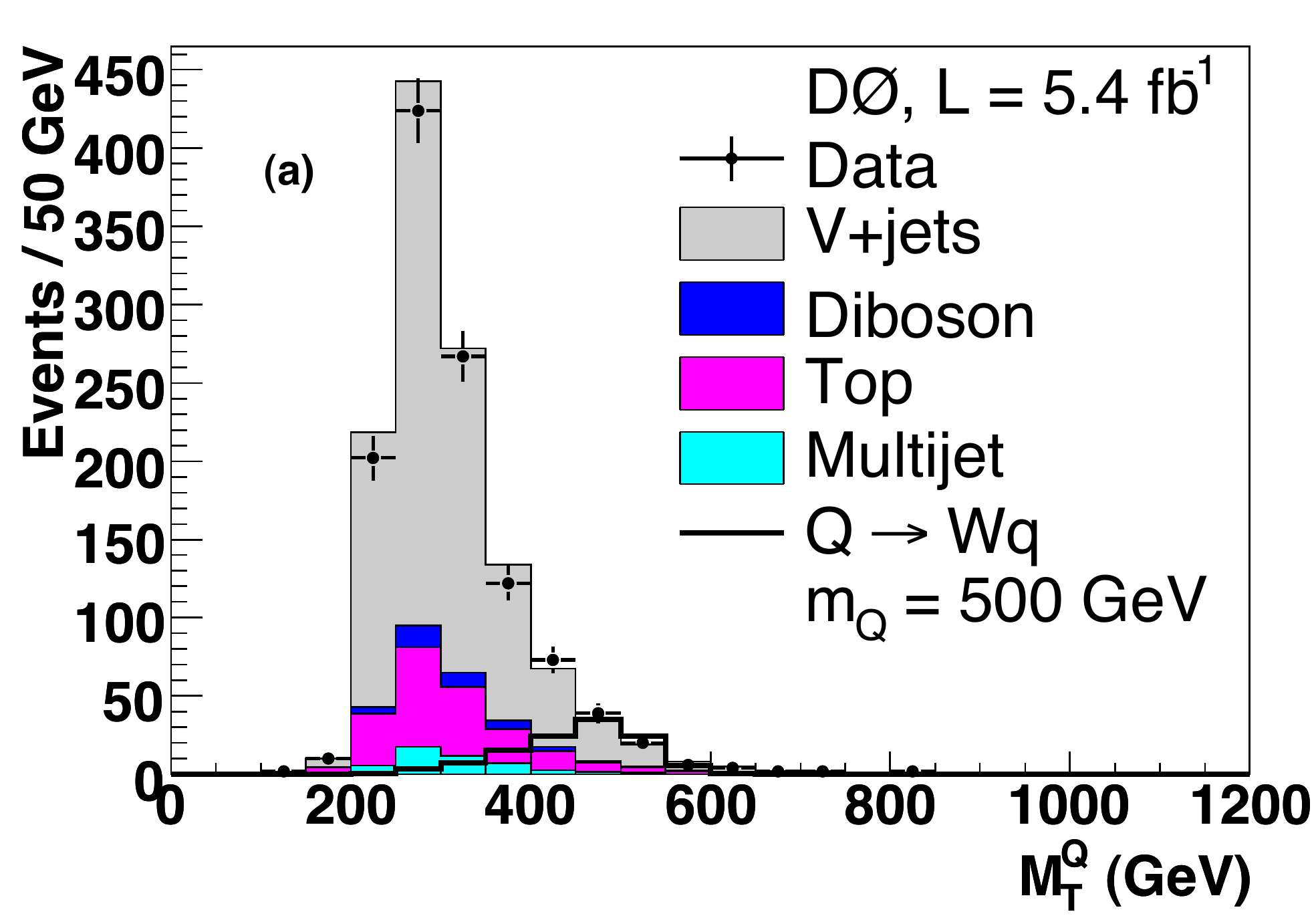}
\includegraphics[width=0.49\textwidth]{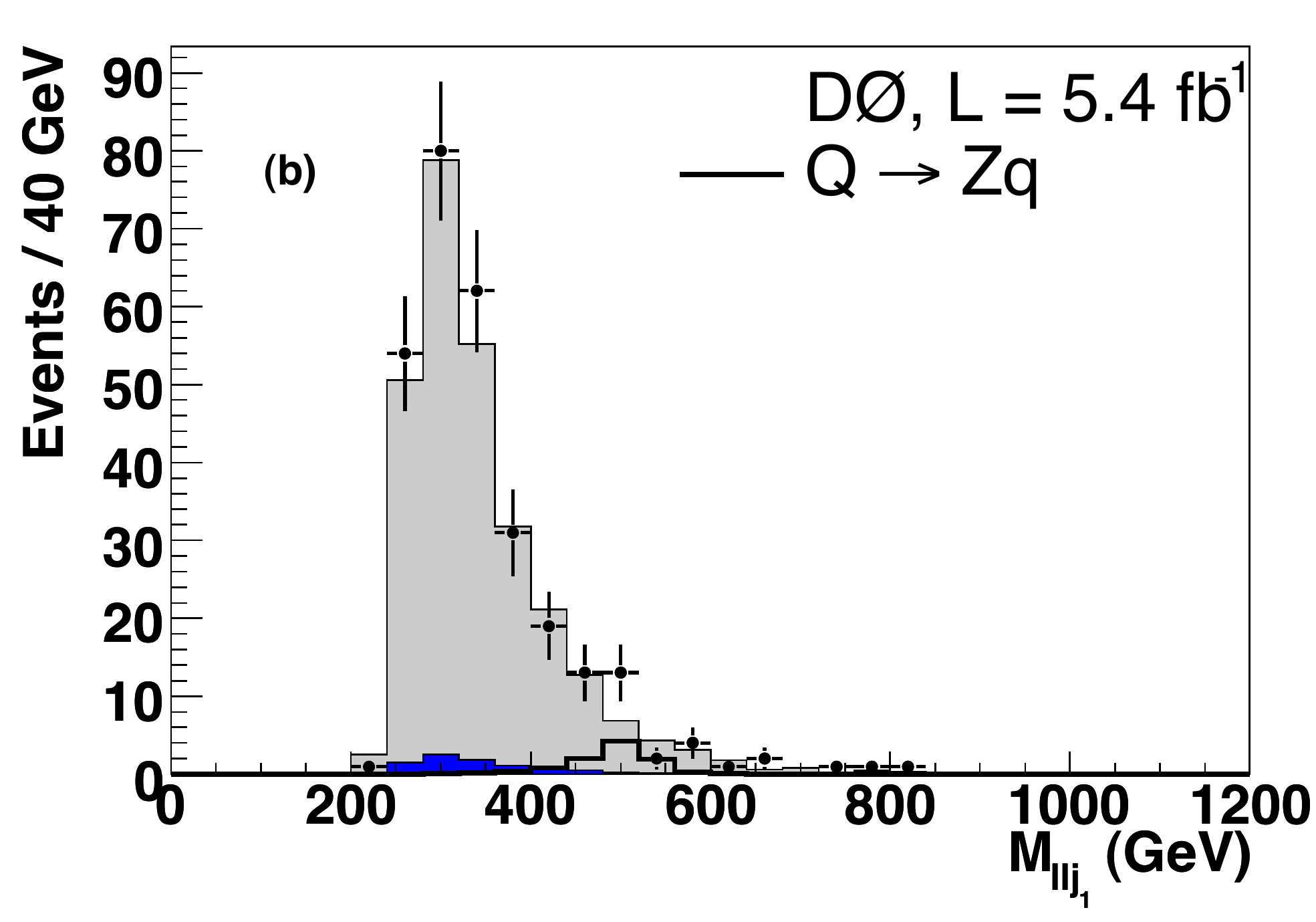}
\caption{{\bf LEFT:} Vector-like quark transverse mass for the single lepton channel. {\bf RIGHT:} Vector-like quark mass for the dilepton channel (right).}
\label{fig:N10IF02}
\end{figure}

\begin{figure}[htbp]
\centering
\includegraphics[width=0.49\textwidth]{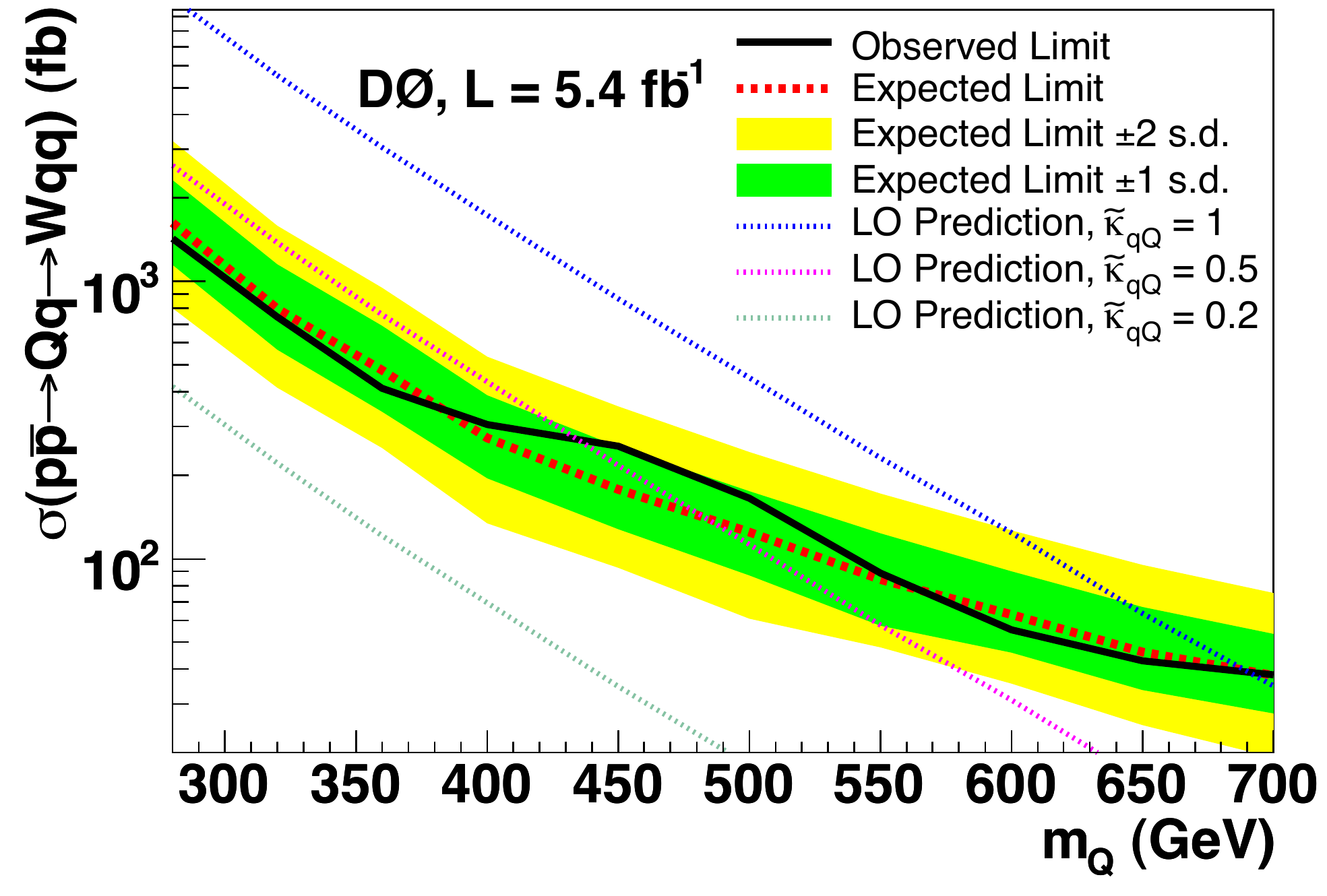}
\includegraphics[width=0.49\textwidth]{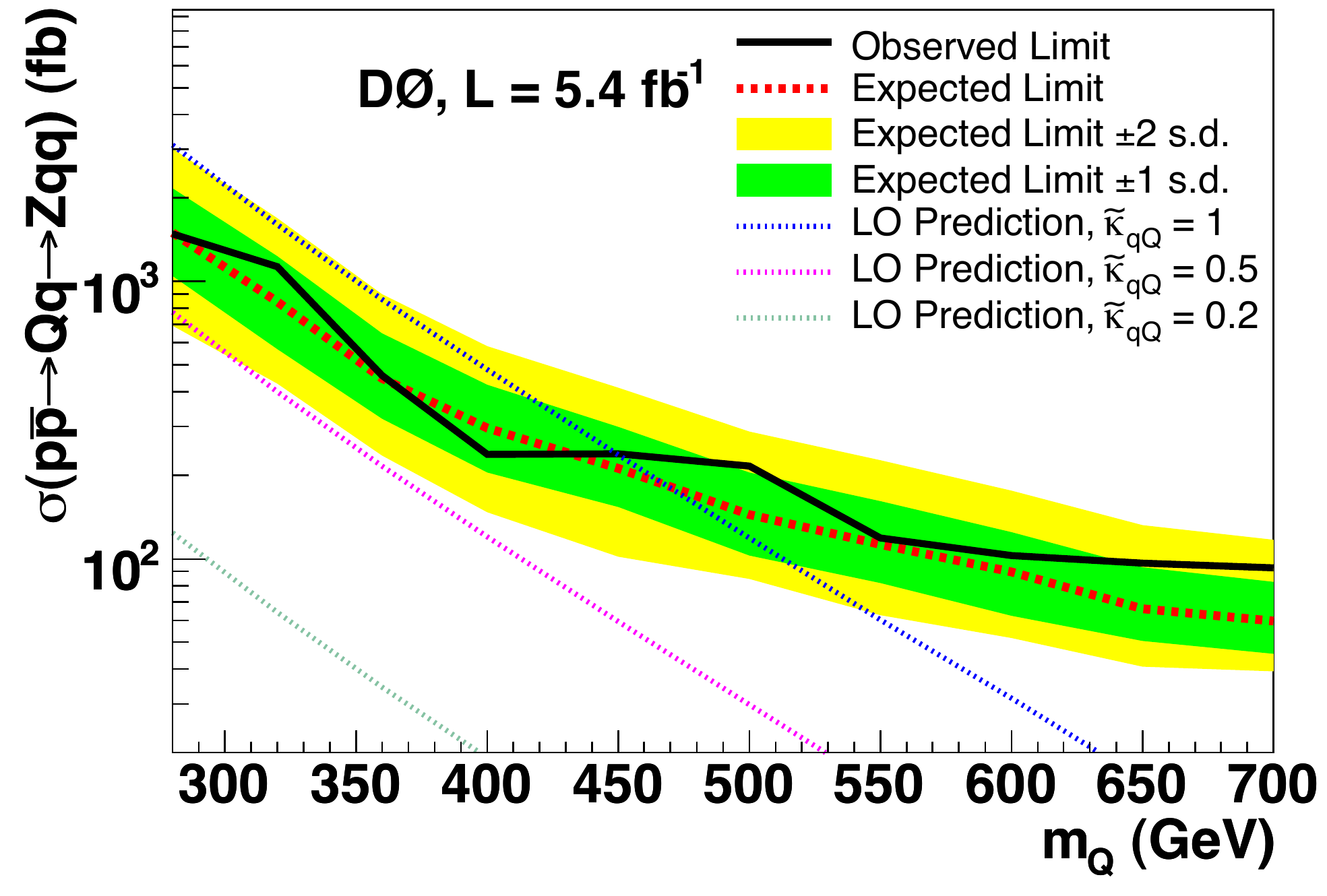}
\caption{D0 observed and expected upper limits on production cross section for a vector-like quark $Q$ decaying to $W+$ jet (left) and $Z+$ jet (right), as a function of $m_Q$ for different values of couplings $\kappa_{qQ}$. }
\label{fig:N10IF03}
\end{figure}

Fig.~\ref{fig:N10IF02} shows the reconstructed vector-like transverse mass for the single lepton channel (left), and the vector-like quark mass in the dilepton channel, reconstructed as the invariant mass of the dilepton + leading jet system (right). 
No significant excess of data over the background prediction in either channel is observed. Upper limits on vector-like quark production 
cross sections are extracted using a modified frequentist approach ($CL_S$)~\cite{cls}, and are shown in Fig.~\ref{fig:N10IF03}.

\begin{figure}[htbp]
\centering
\includegraphics[width=0.6\textwidth]{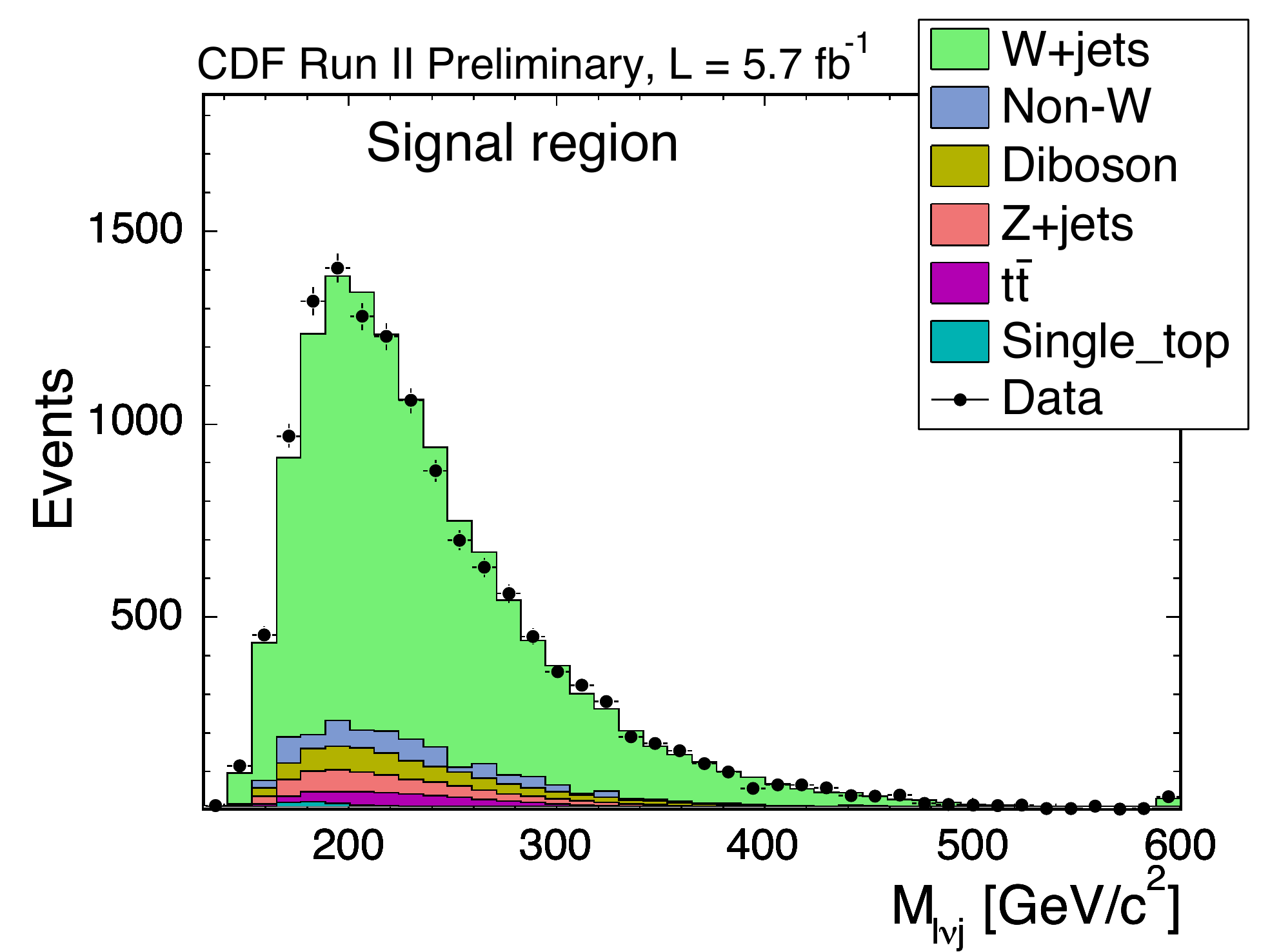}
\caption{Vector-like quark transverse mass in CDF search.}
\label{fig:Mlvj}
\end{figure}

\begin{figure}[htbp]
\centering
\includegraphics[width=0.49\textwidth]{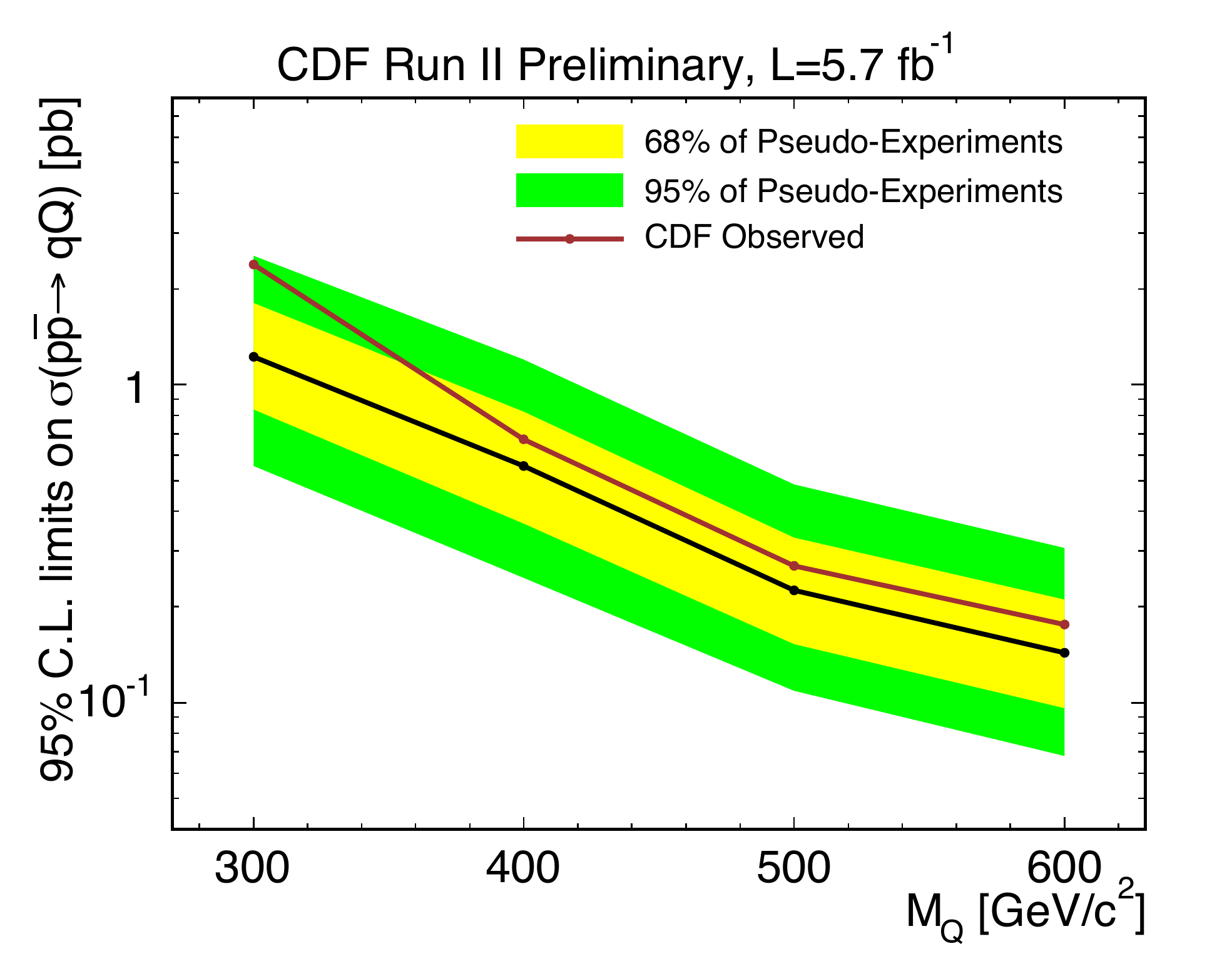}
\includegraphics[width=0.49\textwidth]{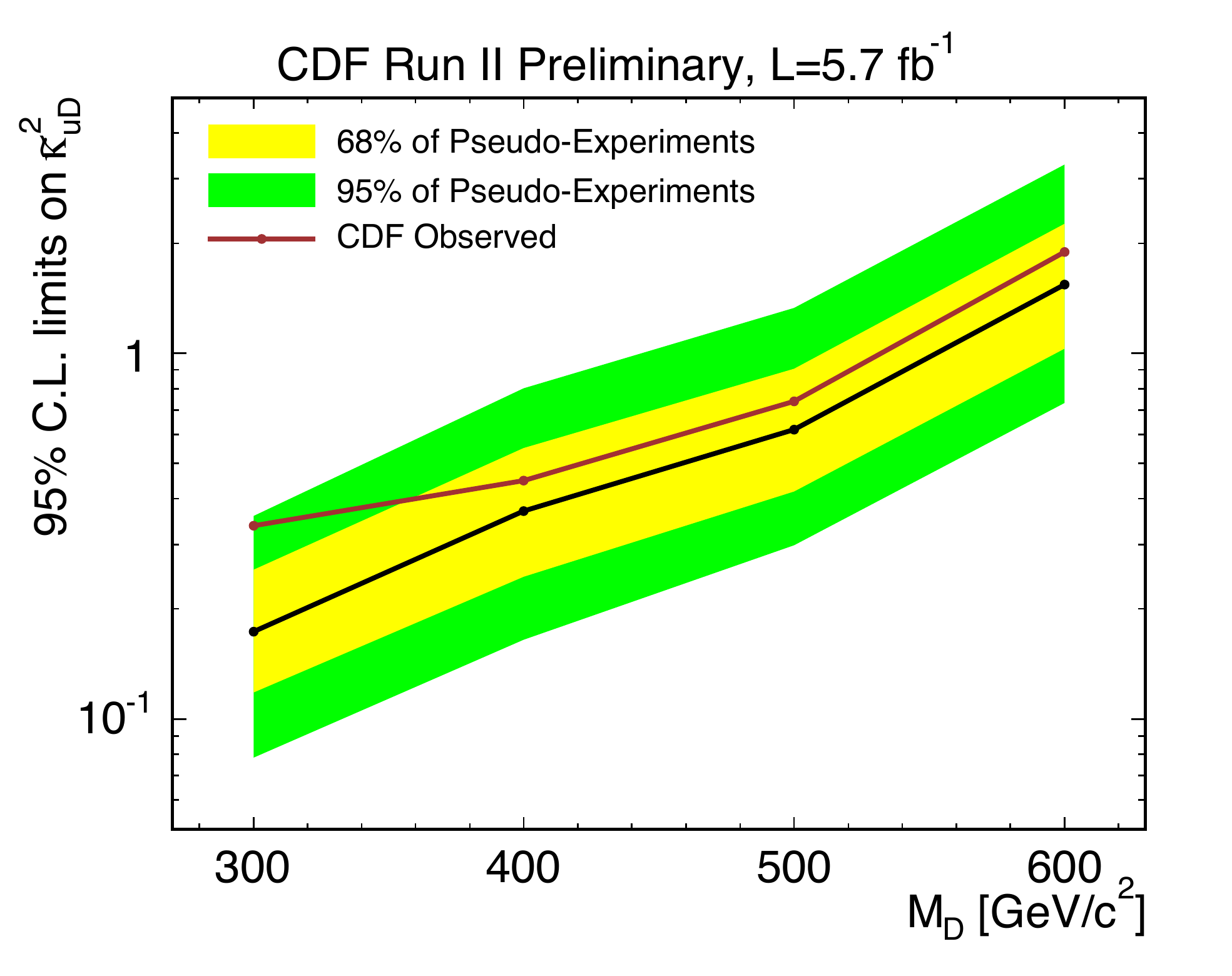}
\caption{CDF 95\% C.L. limit on production cross section of new heavy quarks (assuming $\kappa_{uQ} = 1$) , and 
the limit on the size of the coupling as functions of the heavy quark mass.}
\label{fig:xs_sin}
\end{figure}

CDF Collaboration performed a similar search for singly-produced vector-like quarks decaying to $Wq$ using 5.7 fb$^{-1}$ of data.
The distribution of transverse mass of the vector-like quark is presented in Fig.~\ref{fig:Mlvj}. No significant deviation from the SM 
predictions is observed and the limits on the production cross section of new heavy quarks and the size of the couplings are set, 
that are shown in Fig.~\ref{fig:xs_sin}.

\section{Search for Exotic $T'\bar{T'}$ with $T' \to tX$ }

In some of the exotic fourth-generation scenarios $T'$ quark can decay to the top quark $t$ and a dark matter candidate~\cite{feng}.
The pair production of such quarks  
results in a signature $t\bar{t} XX$ with an extra missing energy due to both of $X$.
CDF performed the search in 4.8 fb$^{-1}$ of data using events $\ell + 4 $ jets $+ \met$, requiring a large missing transverse energy $\met > 100 $ GeV.

\begin{figure}[htbp]
\centering
\includegraphics[width=0.99\textwidth]{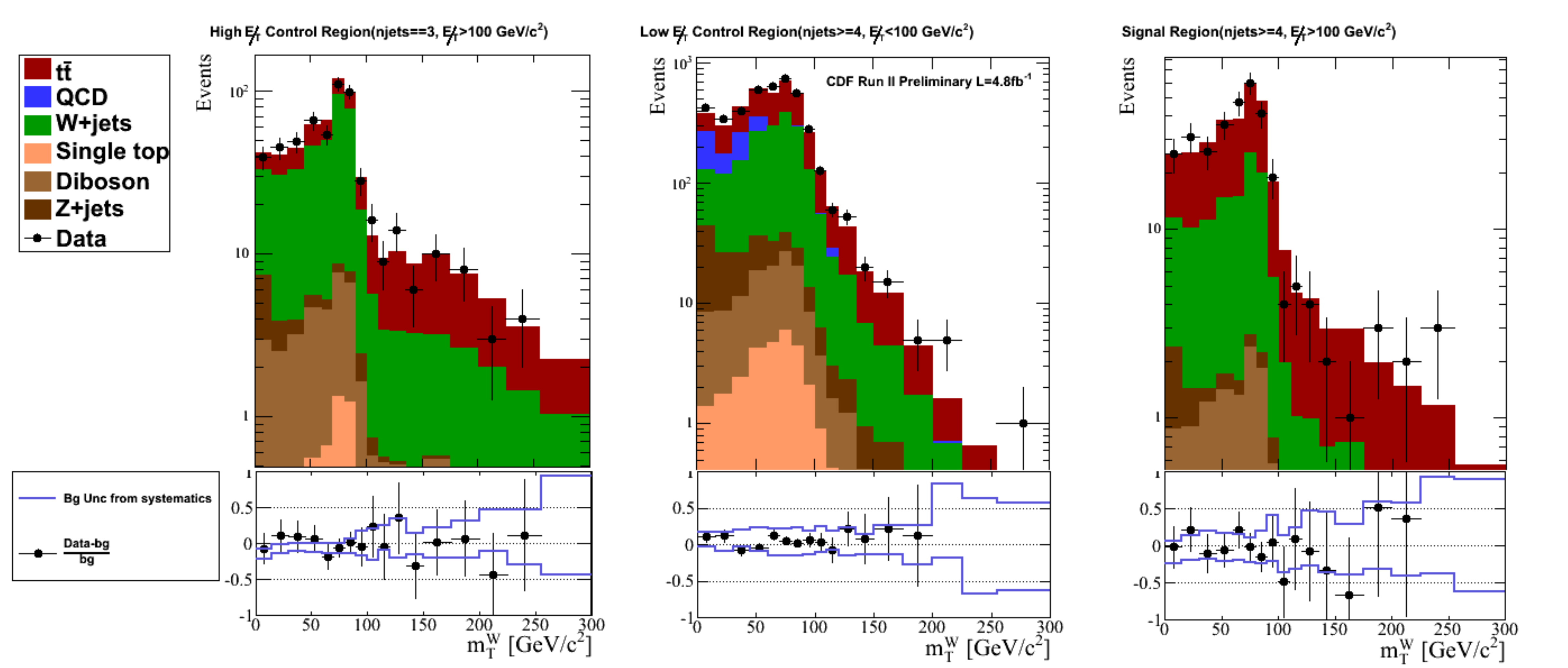}
\caption{Transverse $W$ mass distribution in the control and the signal regions. }
\label{fig:PBmtw}
\end{figure}

The SM background due $t\bar{t}$ and $W+$ jets is validated in two control regions. 
High $\met$ control region: events with $N_{jets} = 3, \met > 100 $ GeV, which validates modeling of the large $\met$ events, 
and low $\met$ control region: events with $N_{jets} \geq 4, \met < 100 $ GeV, which validates the modeling of events with 4 jets.
The signal region is defined as $N_{jets} \geq 4$, and $\met > 100$ GeV.
The transverse $W$ mass distributions corresponding to these regions are shown in Fig.~\ref{fig:PBmtw}. 
The event yields corresponding to different regions are listed in Table~I.

\begin{figure}[htbp]
\centering
\includegraphics[width=0.6\textwidth]{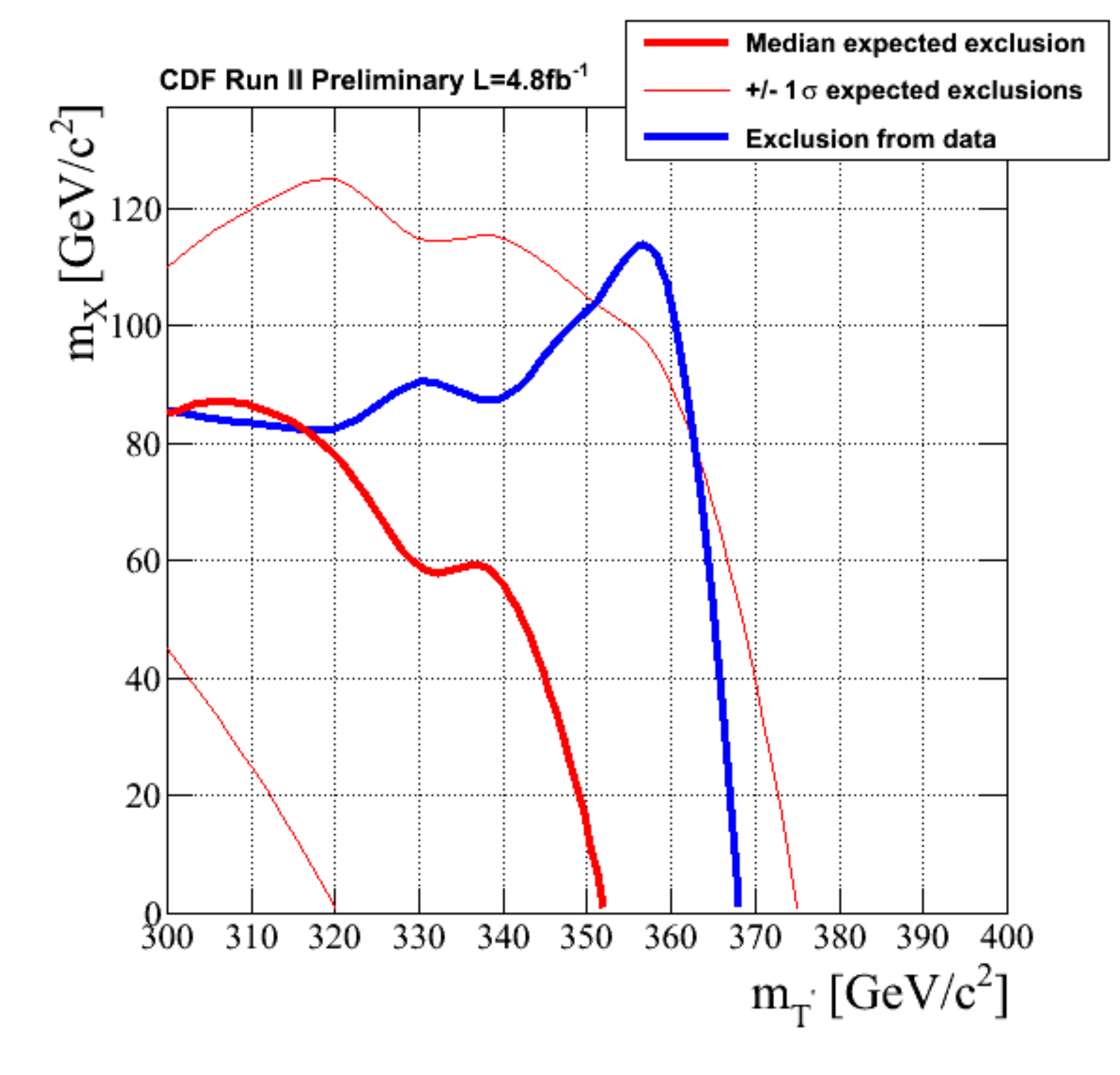}
\caption{Excluded region on the 2D- ($m_{T'}, m_X$) - plane. }
\label{fig:excl}
\end{figure}

No significant excess that could be attributed to production of $T'$ quarks is observed, and the limits are obtained by performing 
a log-likelihood fit to the transverse $W$ boson mass. The observed and expected 95\% C.L. limits as a function of the mass of $T'$ quark and mass of dark matter candidate $X$ are shown in Fig.~\ref{fig:excl}.

\begin{figure}[htbp]
\centering
\includegraphics[width=0.99\textwidth]{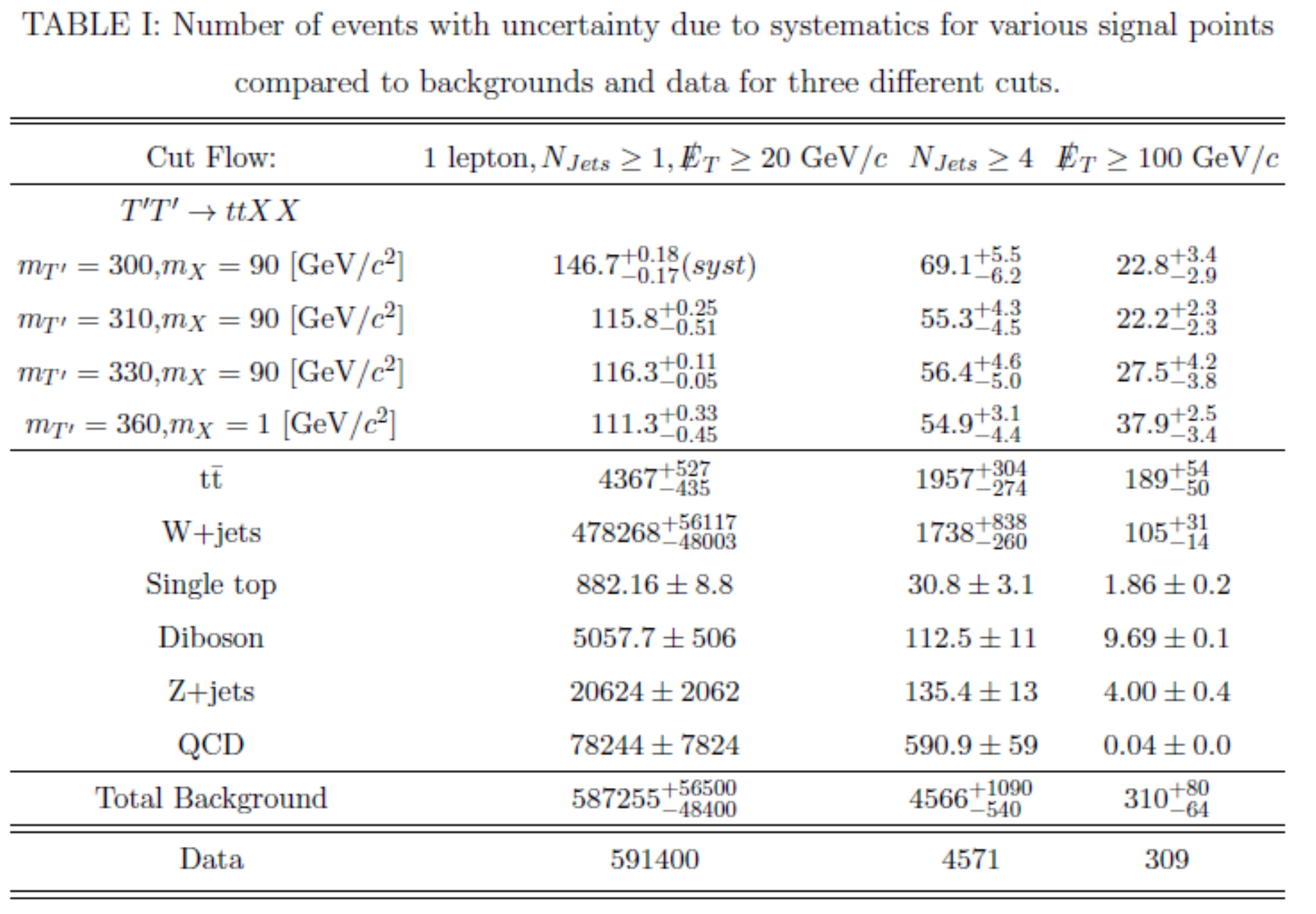}
\end{figure}

\section{Summary }

We presented recent searches for fourth-generation quarks performed by CDF and D0 Collaborations 
using $p\bar{p}$ collisions at $\sqrt{s} = 1.96$ TeV. No evidence for fourth-generation fermions has been 
observed and 95\% C.L. limits on masses and production cross section of these quarks have been set.





\end{document}

%% file: econfmacros.tex
\def\beq{\begin{equation}}
\def\eeq#1{\label{#1}\end{equation}}
\def\eeqn{\end{equation}}

\def\beqa{\begin{eqnarray}}
\def\eeqa#1{\label{#1}\end{eqnarray}}
\def\eeqan{\end{eqnarray}}

\let\bar=\overbar

\def\Dslash{\not{\hbox{\kern-4pt $D$}}}
\def\dslash{\not{\hbox{\kern-2pt $\del$}}}

\def\msb{{\bar{\ssstyle M \kern -1pt S}}}

\newcommand{\met}{\,/\!\!\!\!E_{T}}

\newcommand{\ttbar}{{t\bar{t}}}

\newcommand{\tprime}{t^\prime}

\newcommand{\bprime}{b^\prime}

\newcommand{\ppbar}{p\bar{p}}